\documentclass{article}
\usepackage[utf8]{inputenc}
\usepackage{centernot}
\usepackage{arxiv}
\usepackage[T1]{fontenc}    % use 8-bit T1 fonts
\usepackage{url}            % simple URL typesetting
\usepackage{booktabs}       % professional-quality tables
\usepackage{amsfonts}       % blackboard math symbols
\usepackage{nicefrac}       % compact symbols for 1/2, etc.
\usepackage{microtype}      % microtypography
\usepackage{lipsum}
\usepackage{graphicx}
\usepackage{verbatim}
\usepackage{latexsym}
\usepackage{subcaption}
\usepackage{setspace}
\usepackage{float}
\usepackage{amsmath}
\usepackage{braket}
\usepackage{mathtools}
\usepackage{tensor}
\usepackage[]{algorithm2e}
\usepackage{xcolor}
\usepackage[normalem]{ulem}
\usepackage{wrapfig,lipsum}
\usepackage{multicol}
\title{Modelling equity transactions as bursty processes}
\author{Isobel Seabrook, Paolo Barucca, Fabio Caccioli }
\date{November 2020}
\begin{document}
 
\maketitle
\begin{abstract}
%Trade executions for major stocks come in bursts of activity, which can be partly attributed to the presence of self- and mutual excitation endogenous to the system. In this paper, we study transaction reports for five FTSE 100 stocks, which we model  using Hawkes processes at the overall transaction level, at the level of individual counterparty relationships, and for trades executed via central clearing counterparties to generate synthetic transaction sequences which display similar properties to real transaction sequences. We use a parametric approach to fit the parameters of the Hawkes models to real data, and we suggest methods of choosing participant pairs to transact for models which do not inherently provide this information. We find that the frequency of transactions between counterparties increases the likelihood of them to transact in the future, and that univariate and multivariate approaches at edge level show promise as generative models which are able to produce the bursty, hub dominated systems that we observe in the real world. We further show that Hawkes processes perform well when used to model buys and sells through a central clearing counterparty when considered as a bivariate process, but not when these are modelled as individual univariate processes, indicating that mutual excitation between buys and sells is present in these markets. This study suggests several further explorations of these methods, and applications of generative processes for transaction sequences that would be useful in regulation and industry alike. 

Trade executions for major stocks come in bursts of activity, which can be partly attributed to the presence of self- and mutual excitations endogenous to the system. In this paper, we study transaction reports for five FTSE 100 stocks. We model the dynamic of transactions between counterparties using both univariate and multivariate Hawkes processes, which we fit to the data using a parametric approach. We find that the frequency of transactions between counterparties increases the likelihood of them to transact in the future, and that univariate and multivariate Hawkes processes show promise as generative models able to reproduce the bursty, hub dominated systems that we observe in the real world. We further show that Hawkes processes perform well when used to model buys and sells through a central clearing counterparty when considered as a bivariate process, but not when these are modelled as individual univariate processes, indicating that mutual excitation between buys and sells is present in these markets.
\end{abstract}
\section{Introduction}
Stock trading plays an important role in the economy, providing finance for companies so they can grow and giving investors the opportunity to have a share in the profits of publicly-traded companies. From a regulatory perspective, stock markets require constant monitoring, both from a prudential perspective, as well as from a conduct perspective, to avoid the harms of market manipulation, market abuse, and money laundering. 

%In order to understand how stock markets behave at the microscopic level, and how transactions made by an individual may affect the system as a whole, we need to consider the generative processes 
%Under MIFID II regulations, transaction level data for trades of MIFID II regulated financial instruments is reported to the relevant regulatory bodies. However, the usage and sharing of this data is highly restricted due to its sensitive contents. This means that the generation of synthetic transaction level data is of great importance as it allows financial institutions and regulators to find solutions for technical problems through the sharing of data, to train advanced Machine Learning models and provides opportunities for insights to be gained by the research community into the generative processes underpinning these systems. Regulators have already begun to seek solutions for this, for example through techsprints to collaborate across regulated firms, start-ups, academia, professional services to develop high quality, synthetic financial datasets \cite{fca_techsprint}. Synthetic financial data generation has also been the focus of financial services firms; a detailed discussion can be found in Assefa et. al. \cite{assefa_2019}. 

Under MIFID II regulations, transaction level data for trades of MIFID II regulated financial instruments is reported to the relevant regulatory bodies. However, the usage and sharing of this data is highly restricted due to its sensitive contents. This means that the generation of synthetic transaction level data with realistic properties is of great importance as these could be shared as a surrogate for real data, allowing for instance to outsource the development and training of advanced machine learning models. Regulators have already begun to seek solutions for this, for example through running hackathons and techsprints to collaborate across regulated firms, start-ups, academia, and professional services to develop high quality, synthetic financial datasets \cite{fca_techsprint}. Synthetic financial data generation has also been the focus of financial services firms; a detailed discussion can be found in Assefa et. al. \cite{assefa_2019}.

Market data has received significant interest from the academic community, mainly focusing on time series of stock prices due to the availability of this data. Although many techniques have been developed in an attempt to infer the generative processes that result in the observed prices \cite{turiel, cont2001, nayak2022, todd2016}, due to data sharing constraints there has been less focus on the microscopic behaviors of financial systems through the study of the transactions themselves. The temporal dimension has been studied by observing the arrival times of orders in limit order books, electronic records of the outstanding orders in individual stocks, as this gives a view of the supply and demand in the market. However, since they lack counterparty information, these studies do not take into consideration how the relationships between market participants give rise to the temporal dynamics observed. Our research is unique in its consideration of cross sectional properties observed in the counterparty information of transactions, in combination with temporal properties observed through the transaction timestamps. 

Transaction times are an example of a temporal point process, composed of a time series of binary events that occur in continuous time. Temporal point processes have been used extensively in modelling financial transactions, for example Barcy et. al. \cite{bacry2015hawkes} who model arrival times of buy and sell orders. Temporal point processes have been studied extensively, both following a parametric approach as in this paper, and also non-parametrically, using for instance the approach recently proposed in Dalmasso et. al. \cite{dalmasso_2021}, which makes use of a Variational Autoencoder, applying convolutions to transaction times across interacting sellers and buyers and making use of a recurrent model to encode the sequential structure. Whilst their approach is able to capture some properties of the real transaction network, they were unable to reproduce the overall degree distributions, and, as a black box method, they are unable to draw any further insights from the model parameters or performance. In contrast, we consider fully parametric approaches in this paper, and we discuss an interpretation of the parameter values themselves in providing information about the underlying behaviours of transaction sequences. 

The Poisson process is the most commonly used point process, with applications across finance \cite{Ilalan_2015, Kreinin_2016, scalas_2006} and other applications, in particular in combination with network methods \cite{Rajaram_2005, brown_1994, Houssou_2019, feng_2016, linderman_2014}. However, Poisson processes do not allow for any memory within the process, and are not able to reproduce many of the properties we observe in this paper for transaction sequences, such as burstiness - the tendency for transactions to arrive in bursts of activity. For this study, we consider Hawkes processes \cite{alan_g_hawkes}, which offer additional flexibility to capture these properties as they are based on a counting process in which the intensity function depends on all previously occurred events. Given that burstiness of stock transactions has not been previously studied, we first provide evidence of burstiness for five FTSE100 instruments before demonstrating the utility of the Hawkes process for modelling these systems. 

Hawkes processes have been extensively applied in finance (for detailed reviews of a large range of financial applications, see \cite{bacry2015hawkes, hawkes_2018}), although there are relatively few examples in the literature where the Hawkes process is combined with network models. One such example is presented by Linderman et. al. \cite{linderman2014discovering}, who developed a probabilistic model that combines the mutually-exciting point process with random graph models to produce an implicit network for use cases such as trades on a stock market which are likely to cause subsequent activity on stocks in related industries. They make use of Markov Chain Monte Carlo methods for inference of the resulting model. They apply their methods to discover the latent structure underlying financial markets, which is of use to reveal interpretable patterns of interaction and to provide insight into the stability of markets. Our work differs from their approach as we consider a univariate Hawkes process in which edges are chosen to transact according to some attribute. 
Another relevant application of Hawkes Processes is in the modelling of Limit Order Books, for which there are several examples in the literature \cite{LARGE2007, rambaldi_2016,cartea_2018, xiaofei_2018}. One such example was presented recently in Morariu-Patrichi et. al. \cite{Morariu-Patrichi_2021}, who propose an extension to the Hawkes process to allow them to account for the state of the Limit Order Book, meaning that the effects of price, volumes, bid-ask spread or other properties of the order book have influence on the arrival rate of orders. Following a similar approach for the trade executions themselves would be an interesting next step for our research. For this study, we focus on the use of the original Hawkes process for simplicity and interpretability. 
Given their utility in modelling processes in which events trigger future events, Hawkes processes are also prevalent in non-financial applications, for example in disease modelling \cite{Unwin_2021, Choi_2015} and social networks \cite{zipkin_2016, Pinto_2015}. 

In this paper, since we are not just considering transaction sequences in a univariate sense, but are also exploring the inter-relations between market participants, we make observations of both the burstiness of transaction sequences to evaluate the temporal dimension of transaction sequences, as well as several cross-sectional properties of the systems considered and look to reproduce these with our methods. Properties such as the degree distribution have been a core focus of research in financial networks \cite{BOGINSKI2005431, ENGEL2019, caccioli_2012} which often display fat tailed distributions due to the large range of different market participant types. Higher order properties, capturing the relationships between nodes, have also received attention, such as reciprocity \cite{ENGEL2019, johnson_2015} and assortativity \cite{hurd_2017, Fricke2013},  as well as the rich club coefficient \cite{tao_2022, Alstott2014} which measures the extent to which well connected nodes connect to each other. Assessing our methods against all of these properties allows us to assess their capability to reproduce the behaviours of these systems that are driven by the relationships between market participants.

The main results of this paper are as follows: 
\begin{enumerate}
    \item FTSE100 transaction sequences display bursty characteristics with an observed temporal stability.
    \item Frequency based selection of counterparties to transact enables us to reproduce key cross-sectional properties of the datasets considered.
    \item Models which produce transaction sequences specific for each edge -- which we will refer to as ``edge level'' from this point onwards -- show a stronger performance than those where a single Hawkes process is used to generate events independently of the pair of counterparties involved. We will refer to the latter as ``overall network level'' in the following.
    \item Multivariate Hawkes processes which capture mutual excitation between pairs of counterparties show stronger performance in reproducing cross-sectional properties of the systems than those which only allow self-excitation.
    \item Strong performance is observed when applying a model which captures mutual excitation between buy and sell sequences to trades executed via a single hub.
\end{enumerate}

\section{Methods}
When considering transaction networks, there are a number of different excitation mechanisms which may result in a trade. There may be some level of mutual excitation, in which the presence of a trade between two individuals excites future trades between other pairs of individuals, and there may be excitation on the level of an individual pair of traders, where the presence of a trade between them excites (or inhibits) the presence of a trade between them in the future. There will also be trades that are controlled by exogenous factors, such as news events. Further to this, we can also not rule out the possibility of an inherent excitation in the transaction sequence as a whole, regardless of who the participants are. This is supported by our observations in section \ref{sec:data exploration}, where we see for all datasets considered a consistent level of burstiness for transaction sequences across all participants.
In order to take us a step closer to understanding the generative process underlying these networks, we compare a number of different approaches using the Hawkes process to simulate transactions. We compare the use of both univariate (at the edge level and overall network level) and multivariate Hawkes processes, and compare to both univariate and multivariate Poisson processes as null models. 

For the univariate case at the overall network level, we are fitting the Hawkes process to the transaction sequence, disregarding the information on which edge is transacting. This means that for these methods, we also need a method to select which edge will transact at each time period. 
Given our observations in a previous study of the predictability of subsequent transactions given the importance of nodes and edges in snapshot networks of similar equity networks \cite{seabrook2020}, we first consider whether the spectral measure of edge importance we propose can be used in a weighted selection to choose which participants should transact. This measure of importance captures the sensitivity of the network's leading eigenvalue $\lambda$ to changes in an individual edge $A_{ij}$, and can be approximated as: 
\begin{equation}
    \frac{\partial\lambda}{\partial A_{ij}}=2x_{0,i}x_{0,j}
\end{equation}
Where $x_{0,i}$ and $x_{0,j}$ are the $i$th and $j$th components of the eigenvector corresponding to the leading eigenvalue. We also compare this to selecting the edge for each transaction according to the frequency of their historical transactions, and also to simply selecting edges randomly as a null model.

For the multivariate case, the Hawkes process inherently produces the information on the participants transacting with each other, so we do not need to make use of the above approach for selecting the participants for each transaction. 
%Since our measure of importance, being spectral based, is designed to capture an edge's influence on the global network activity, we hypothesise that importance might be a useful proxy in capturing the cross-sectional behaviours of these networks, providing a scalable model for temporal evolution of bursty networks. 

% Overall, this means we are comparing six different methods of generating bursty transaction sequences, of which three require further methods to select which edges will change, for which we consider using three methods to do so, meaning that overall we are comparing 15 different models for generating bursty transaction networks. We will now give further details of the constituent parts. 

\subsection{Transaction sequence generation methods}
\subsubsection{Poisson processes}
The Poisson process is one of the most widely used point processes, used in scenarios where events occur at a constant rate. Formally, the Poisson process is a renewal process in which the interarrival times have an exponential distribution, with a density of interarrival times of 

\begin{equation}
  f_X(x) = \lambda e^{-\lambda x} 
\end{equation}
 for $x>0$, and here  $X$ is the set of interarrival times and $\lambda$ refers to the arrival rate of the process \cite{poisson}. A key feature of the Poisson process is that it is memoryless, which is a disadvantage when considering financial transactions, in which the market responds to activity. 
\subsubsection{Hawkes processes}
Hawkes processes are point processes defined by the following conditional intensity function:
\begin{equation}
    \lambda_i(t)=\mu_i+\sum_{j=1}^D\sum_{t_k^j<t}\phi_{ij}(t-t_k^j)
\end{equation}
where $\mu_i$ is the background rate of the process $i$, sometimes referred to as the baseline, $t_k^j$ are the timestamps of all events of node $j$, and $\phi_{ij}$ are the kernels - functions which govern the clustering density of the point process, and are sometimes also referred to as excitation functions. In the literature, the Hawkes kernels usually take the form of an exponential or power law \cite{ZHANG2016762}, and in this paper, we consider exponential excitation functions, so $\phi_{ij}$ takes the form 
\begin{equation}
    \phi_{ij}(t) = \alpha^{ij}\beta e^{-\beta t} 1_{t>0}
\end{equation}
where $\alpha^{ij}$ is the kernel intensity governing the extent to which excitatory behaviour dominantes over the background process, and $\beta$ is the kernel decay governing the timescale over which a transaction influences future transactions. The baseline parameter $\mu_i$ can either be constant or can be time varying in nature. While estimation of the process in the case of a time varying baseline has been tackled by Chen et. al. \cite{Chen_2013}, who take a non-parametric approach to estimate the process, for simplicity and following observations of stationarity in the properties of the process, we opt to consider a constant baseline in this study. 

When using the Hawkes process as a univariate model for a single transaction sequence, $i=1$ and the Hawkes process reduces to 
\begin{equation}
    \lambda(t)=\mu +\sum_{t_k<t} \phi(t-t_k)
\end{equation}
Written as a probabilistic model which defines the probability of a transaction at time $t$ given the history of the process up to time $t$, $\mathcal{H}_t$, 
\begin{equation}
    \mathcal{P}(N(t+h)-N(t)=1|\mathcal{H}_t) = \lambda(t|\mathcal{H}_t)h+o(h)
\end{equation}
where $h$ is an infinitesimally small number \cite{laub}. The history $\mathcal{H}_t$ is the time-ordered sequence of previous events.  

Table \ref{tab:hawkes_summaries} outlines the different ways we make use of the Hawkes process and the dataset subsets we apply these to. 
\begin{table}[]
    \centering
    \begin{tabular}{|c|c|c|}
        \hline
     Hawkes type & Granularity & Dataset \\
     \hline
    Univariate & Overall transaction sequence & All venues \\
    & & Single venue \\
    & & Off venue \\
    Univariate & Edge level & All venues\\
    & & Single venue \\
    & & Off venue \\
    Multivariate & Edge level & All venues\\
    & & Single venue \\
    & & Off venue \\
    Univariate & Buys/sells & Single CCP \\
    Bivariate & Buys/sells & Single CCP \\
    \hline
    \end{tabular}
    \caption{Summary of different applications of Hawkes Processes considered in this paper}
    \label{tab:hawkes_summaries}
\end{table}

\subsubsection{Parameter estimation}
%For the case of a univariate Hawkes process, it is controlled by three parameters which can be estimated using Maximum Likelihood Estimation. In the univariate case, the log-likelihood can be expressed as \cite{zhou13}
A univariate Hawkes process is controlled by three parameters, which can be estimated using Maximum Likelihood Estimation. The log-likelihood can be expressed in this case as \cite{zhou13}
\begin{equation}
    \mathcal{L} = \log \frac{\prod_{i=1}^{n}\lambda(t_i)}{\exp{\int_0^T \lambda(t)dt}} = \sum_{i=1}^n \log \lambda(t_i)-\int_0^T \lambda(t)dt
\end{equation}

However, considering the multivariate Hawkes Process with mutual excitation between edges, the number of parameters to be estimated increases as $\mathcal{O}(2n + n^2)$, as each edge will have its own baseline intensity and kernel decay parameters, as well as cross terms within the kernel matrix to capture the influence of each edge level process on every other process. This means that even for a modest number of processes, a standard Maximum Likelihood approach is computationally infeasible \footnote{On a  2nd generation Intel Xeon Platinum 8000 series processor, the parameter estimation for a system with 200 cross terms took over 8 hours to complete.}. Instead, we make use of the ADM4 method developed by Zhou et. al. \cite{zhou13}, using the implementation provided by the tick package\cite{tick}, to estimate the $\mu_i$ and $\alpha_{ij}$ parameters for a process with a given decay parameter $\beta$, and we use a brute force approach to find the value for $\beta$ which produces a sequence with a burstiness closest to the real sequence. 
The ADM4 method first of all makes use of the sparsity and low-rank properties of common network systems, and it applies sparse and low-rank regularisation to the likelihood function. Since the resulting likelihood is non-differentiable and difficult to optimise, it then combines the idea of alternating direction method of multipliers and majorization minimization to convert the optimisation problem to several sub-problems that are much easier to solve. The performance of this method is demonstrated in \cite{zhou13}, in application to both simulation and real-world datasets. A drawback of this method is that it is only implemented with a constant kernel decay across all processes, whereas we know from our exploration later in this paper that the individual processes are highly heterogeneous in terms of their burstiness. Fast estimation of a multivariate Hawkes with the flexibility of a varying kernel decay would thus be a useful area for future research efforts.

\subsection{Edge selection methods}
Following our observations of the role of importance in predicting subsequent presence of edges in equity transaction networks \cite{seabrook2020}, the first method we consider for selecting edges is to select the edge to transact at each timestamp probabilistically, with the edges weighted by the product of the nodes' importance. 
There are a number of limitations to this method - first, a balance needs to be struck between having a model in which the full set of node importances is updated at each timestamp, and runtime, since to recompute all node importances at each timestamp would be computationally expensive for a reasonable length of transaction sequence. In this paper, we make the compromise of updating node importances every 10 transactions, as this was found to have an acceptable runtime. Secondly, this method only allows edges which contain nodes which have previously appeared to be selected to transact, not allowing the inclusion of new nodes. This method could also be used with other definitions of node importance or centrality, which would be an interesting avenue for future work. 

We also consider selection of edges to transact according to their transaction frequency in the real data. This method has a significantly lower computational cost than the importance based selection. However, it still only allows for the inclusion of nodes that existed in the training set. Finally, we compare both methods to randomly selecting the edge to change from the set of edges that exist in the real data. 

\subsection{Model evaluation}
In order to evaluate any models proposed, we need to assess their ability to reproduce the behaviours observed in real networks, both at large and small aggregations, and both aggregate statistics and the more granular individual node or edge behaviours. Several researchers have presented methods for the evaluation of network models, for example Zhang et. al. \cite{zhang_2009} focus on the symbiotic effect, or in other words the extent to which the most important nodes connect to each other, as a criterion for whether or not Barabasi-Albert models and variations on this model, are suitable for the network at hand. Their method quantifies the difference between the degree based and betweenness based rich club curves, and they apply their methods to four widely studied networks demonstrating their ability to detect varying levels of the symbiotic effect. 
%Other examples of research focusing on specific properties of the networks that a good model should be able to reproduce include \footnote{citations to be included}. 
Wang et. al. \cite{wang_2011}, propose an approach making comparisons of likelihoods of the observed network being driven by different models, as a means of finding the best model without relying on pre-selected model features. This method has benefits since it does not rely on selecting specific properties of the system to reproduce, however lacks in the interpretability and doesn't allow any insights to be gained into the interplay between temporal and cross sectional properties of the system.  

% When considering point process models, such as the Hawkes process, a likelihood approach is also one of the most common ways of assessing the model suitability. This is explored by Chen et. al. \cite{Chen_2018} for an extension to the Hawkes process called the Renewal Hawkes process, for which the likelihood was previously thought to be infeasible to evaluate. %MORE. IF INCLUDING POINTS ABOUT LIKELIHOOD MODEL EVALUATION, WE EITHER NEED TO DO IT OR JUSTIFY WHY NOT. 

In evaluating our models, we focus on reproducing the burstiness as a key temporal property of transaction sequences, as this not only allows us to assess the fit of the Hawkes process, but also allows us to gain insights from the burstiness observed and the parameter values that give rise to this. %following the likelihood methods presented in Wang et. al. \cite{wang_2011} (TBC). 
Burstiness can be directly derived from the sequence of inter-transaction times, by comparing to the statistical properties of the inter-transaction times for a Poissonian sequence. Starting from the coefficient of variation of the inter-contact times ($\frac{\sigma_{\tau}}{\mu_{\tau}}$), which equals 1 for a Poissonian sequence, the burstiness can be defined as 
\begin{equation}
    B=\frac{\frac{\sigma_{\tau}}{\mu_{\tau}}-1}{\frac{\sigma_{\tau}}{\mu_{\tau}}+1} = \frac{\sigma_{\tau}-\mu_{\tau}}{\sigma_{\tau}+\mu_{\tau}}
\end{equation}
where $\sigma_\tau$ is the standard deviation of the inter-contact times, and $\mu_tau$ is the mean. $B=1$ indicates a very bursty sequence, $B=0$ a Poissonian sequence and $B=-1$ an entirely periodic sequence \cite{Holme}.

We also need to consider the cross sectional properties that our models produce, either through the direct edge selection methods in the case of the univariate modelling approach, or in terms of the connections that a multivariate Hawkes process brings about. We consider the a range of cross-sectional properties: 
\begin{itemize}
    \item \textbf{Degree distribution} - the fraction of nodes in the network with degree $k$
    \begin{equation}
     P(k)=\frac{n_k}{n}
    \end{equation}
    where $n_k$ is the number of nodes with degree $k$ and $n$ is the total number of nodes
    \item \textbf{Assortativity} - 
    Assortative mixing in networks is the tendency for nodes to connect to other nodes that are like them. In this paper we consider degree assortative mixing, which is quantified in terms of the quantity $e_{ij}$, which is the fraction of edges in a network that connect a vertex of degree $i$ to one of degree $j$. The assortativity coefficient then quantifies the level of assortative mixing, given by: 
    \begin{equation}
        r=\frac{\sum_i e_{ii}-\sum_i a_i b_i}{1-\sum_i a_i b_i} = \frac{\mathrm{Tr} (\mathbf{e}) - ||\mathbf{e}^2||}{1-||\mathbf{e}^2||}
    \end{equation}
    where $\mathbf{e}$ is the matrix whose elements are $e_{ij}$, $a_i = \sum_j e_{ij}$ and $b_i=\sum_j e_{ji}$ are the fraction of each type of end of an edge that is attached to vertices of type $i$, and $||\mathbf{x}||$ means the sum of all elements of the matrix $\mathbf{x}$ \cite{newman_03}. 
    \item \textbf{Reciprocity} - Reciprocity is defined for a directed graph as the ratio of the number of edges pointing in both directions to the total number of edges in the graph $G$
    \begin{equation}
        r = \frac{|(u,v) \in G||(v,u) \in G|}{|(u,v) \in G|}
    \end{equation}
    similarly, for a single node, reciprocity is the ratio of the number of edges in both directions to the total number of edges attached to the node in question \cite{garlaschelli_2004}. 
    \item \textbf{Rich club index} -
    the rich club phenomenon in networks is characterised when the hub nodes with high degrees are on average more intensely connected than the nodes with smaller degree \cite{mcauley_07}. This can be measured using the rich club coefficient at each degree $k$, which is the fraction of of the actual and potential number of edges among the set of nodes with degree higher than $k$: 
    \begin{equation}
        \phi(k) = \frac{2E_{>k}}{N_{>k}(N_{>k}-1)}
    \end{equation}
    where $N_k$ is the number of nodes with degree larger than $k$ and $E_k$ is the number of edges among those nodes. 
\end{itemize}

Here, we are considering higher and higher order properties allowing us to study the cross-sectional behaviour of the systems studied at first the node level through the degree distribution, then the level of neighbours through assortativity and beyond through network level reciprocity and rich club distributions. 
\section{Results}
\subsection{Data exploration}
\label{sec:data exploration}
In this paper, we make use of transaction reports containing the details of 5 different FTSE100 constituent instruments, reported under MIFID II regulations. These datasets were available in their raw transaction form, containing information on price, volume, transaction time and anonymized identities of market participants, and the instruments studied were selected at random. The transactions are reported to the nearest microsecond, providing a highly detailed view of the behaviour of these markets at the lowest level of granularity. Although all of the instruments are FTSE100 stocks which will be dominated by high frequency trading strategies, for the day considered they vary significantly in the number of transactions, with the smallest being 16,000 transactions (on average approximately one transaction every 2 seconds) and the largest 110,000 transactions (on average approximately one transaction every 0.2 seconds). It is also worth noting that when working with these trade reports, we need to account for duplicate reporting, in which more than one counterparty involved in the trade report the same trade. Whilst care has been taken in doing this for this paper by identifying trades occurring at the same time and quantity between matching counterparties, we note that reporting quality issues may result in trades being duplicated with differences. Further work is needed to fully account for this. 

\begin{figure}
    \centering
    \includegraphics[width=\linewidth]{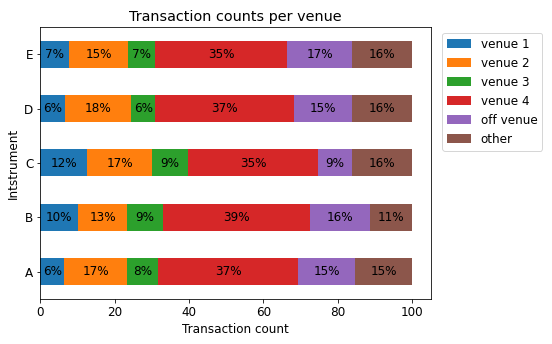}
    \caption{Transaction counts by venue, showing the top 4 venues and off venue trades separately and the remaining venues aggregated (`other').}
    \label{fig:venue_counts}
\end{figure}

As the majority of trading in these liquid instruments is through market exchanges, one of the prevalent properties of these systems is the dominance of central institutions, usually central counterparty clearing houses (CCPs). A CCP is an institution that sits between parties involved in a transaction, to absorb any counterparty credit risk. For the instruments in question, all show $30-40\%$ of trades intermediated by the same central clearing house. For most major venues we see order book trades cleared through a single CCP for each instrument, however with the existence of interoperability agreements which give clearing members the opportunity for netting across CCPs covered by these agreements, it is not uncommon to see trades cleared through several CCPs for the same instrument\footnote{This is not the situation for many other asset classes, for example ICE futures contracts will be cleared through ICE only.}.  Despite this, we do see a large proportion of trades through one venue, as can be seen in figure \ref{fig:venue_counts}. We thus consider application of Hawkes processes in various different ways, to different subsets of these systems: we consider application to all the visible trades for these instruments, trades through a single exchange, trades executed off exchange\footnote{Off exchange trades of exchange listed securities occur for a number of reasons, for example they may occur between different arms of a large institution in different jurisdictions in order to move client money between these different arms}  and trades only with the largest participant, a prominent clearing house. It is theoretically possible to disintermediate the central clearing house, however most of the trading in these cases will be on Central Limit Order Books, the buyers and sellers either side of the clearing house will not deliberately trade with each other, but are matched according to the order matching system in place \cite{Janecek_07}. Also, by trading via the clearing house, the risk of the trade is absorbed by the clearing house. For these reasons, we decide not to disintermediate the central clearing house as it represents a true node in the transaction network. 

Of the trades operated through the major venues, we also see different types of market - for example, the second and third largest venues for all instruments considered in this paper are both examples of dark order book trading venues\footnote{The top four venues are the London Stock Exchange, the LSEG's Turquoise Lit venue, and the BATs Europe and Chi-X Europe venues operated by Cboe Europe.}. These are venues in which there is a complete lack of transparency of the order book, allowing large institutional investors to place trades that do not impact the markets \cite{financial_stability_review}. On the other hand, the fourth largest is a lit order book venue, where the full book of orders is visible to all market participants. We would expect very different trading mechanisms for lit vs. dark order books, as the former allows the market price to respond more readily than the latter.

\begin{figure}
    \centering
    \includegraphics[width=\linewidth]{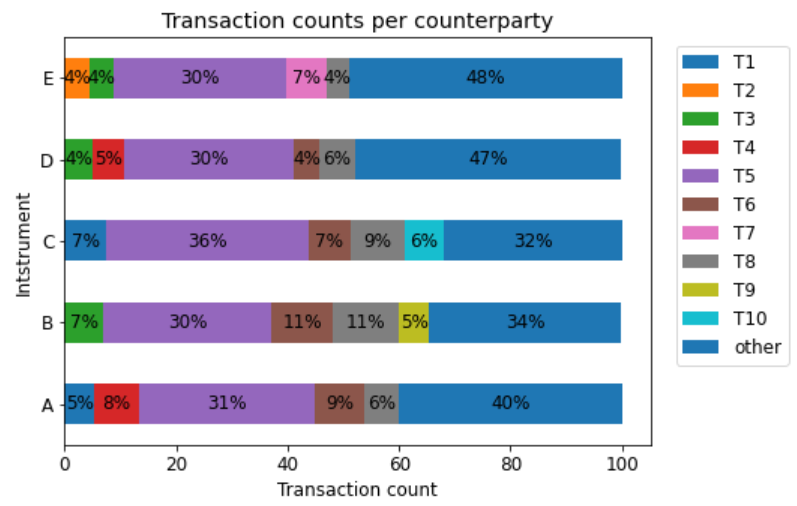}`    \caption{Transaction counts by counterparty, showing the top 5 counterparties for each instrument separately and the remaining counterparties aggregated (`other').}
    \label{fig:counterparty_counts}
\end{figure}

\begin{table}
    \centering
    \begin{tabular}{|c|c|c|c|c|}
    \hline
    Data & All & Single venue & Off venue & Single CCP \\
    \hline
    A & 0.35 & 0.24 & 0.25 & 0.27 \\
    B & 0.48 & 0.32 & 0.23 & 0.39 \\
    C & 0.37 & 0.22 & 0.27 & 0.28 \\
    D & 0.30 & 0.20 & 0.14 & 0.20 \\
    E & 0.46 & 0.33 & 0.18 & 0.35 \\
       \hline
    \end{tabular}
    \vspace{0.1cm}
    \caption{Burstiness of the 5 stocks, considering the entire dataset, only trades on the venue with the most trading activity, only off venue trades and only trades through the main clearing house (CCP).}
    \label{tab:bursts_hub}
\end{table}

\begin{figure}
    \centering
    \includegraphics[width=\linewidth]{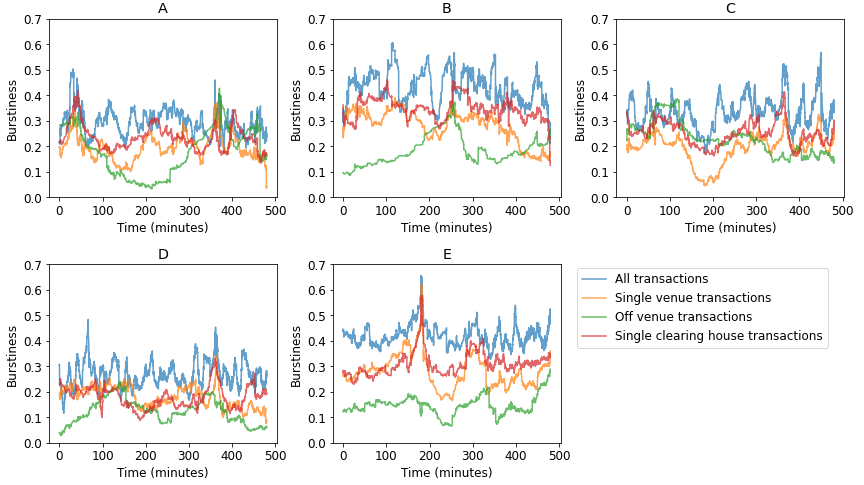}
    \caption{Network level burstiness across rolling windows of length 200 transactions, for the different data subsets.}
    \label{fig:rolling_network_burstiness}
\end{figure}

Considering these subsets for all 5 instruments, the overall network burstinesses are presented later in this paper in table \ref{tab:burst_res}, and in figure \ref{fig:rolling_network_burstiness}, we present the network level burstiness computed across rolling windows containing 200 transactions, with the window size chosen as an optimal size in containing enough data points for each window for all datasets. We can see from these that all instruments have a similar burstiness for the full transaction set, and in general the off venue subset shows the lowest burstiness and in some cases shows a visible dip in the burstiness around midday. If we consider applying an Augmented Dickey-Fuller test for stationarity to the burstiness calculated in a rolling window of 200 transactions, the null hypothesis of a unit root present (which would indicate non-stationarity) is rejected at 5\% significance for all except one of the transaction sequences.

\begin{figure}
    \centering
    \includegraphics[width=\linewidth]{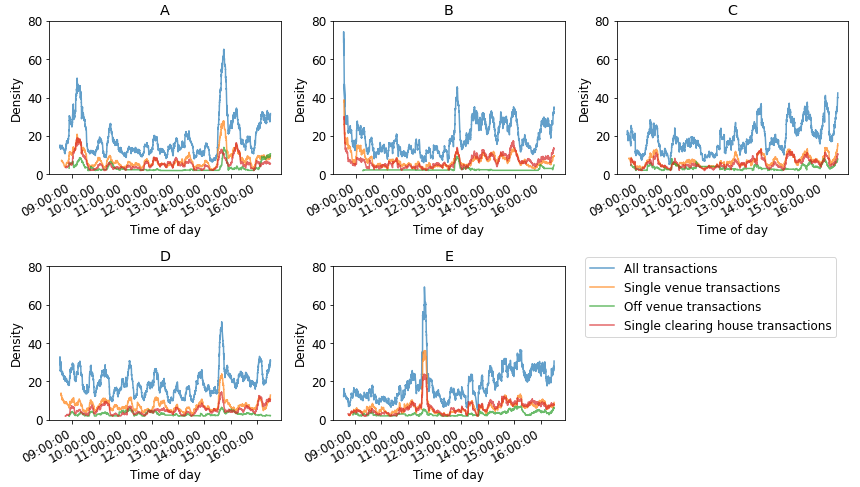}
    \caption{Network level density (number of transactions in a given time window) across rolling 10 minute windows}
    \label{fig:rolling_network_density}
\end{figure}

Figure \ref{fig:rolling_network_density} shows the rolling transaction density (number of transactions in a given time window) for 10 minute time windows, again with the window size chosen optimally to contain enough data points in each window for all datasets. We see that unlike the burstiness, the transaction density is in general not stationary, and in some cases, shows a visible trend. 

\begin{figure}
    \centering
    \includegraphics[width=\linewidth]{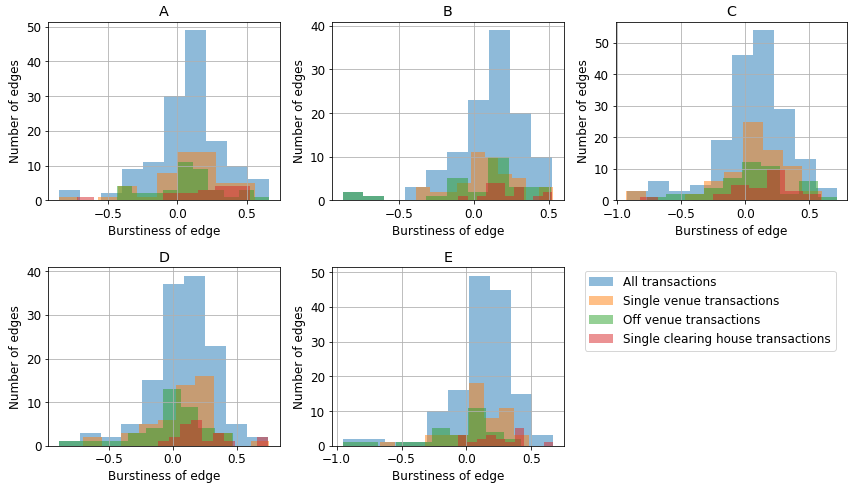}
    \caption{Histograms of the burstiness of individual edges across the entire observation period, for the 5 different datasets for their different subsets}
    \label{fig:edge_burstiness}
\end{figure}

When considering individual edge burstiness, we see that the individual edges tend to show a lower burstiness, with average edge burstinesses between 0.1 and 0.3 for the five datasets, in comparison to the network as a whole which shows average burstinesses between 0.3 and 0.5. This suggests that  that mutual-excitation is a contributing factor to the burstiness. In figure \ref{fig:edge_burstiness} we see a significant variation in the edge level burstiness, with some edges showing periodic behaviour (negative burstiness). The subsets containing only trades with the most active central clearing house show the largest average burstiness across the edges. 

\begin{figure}
    \centering
    \includegraphics[width=\linewidth]{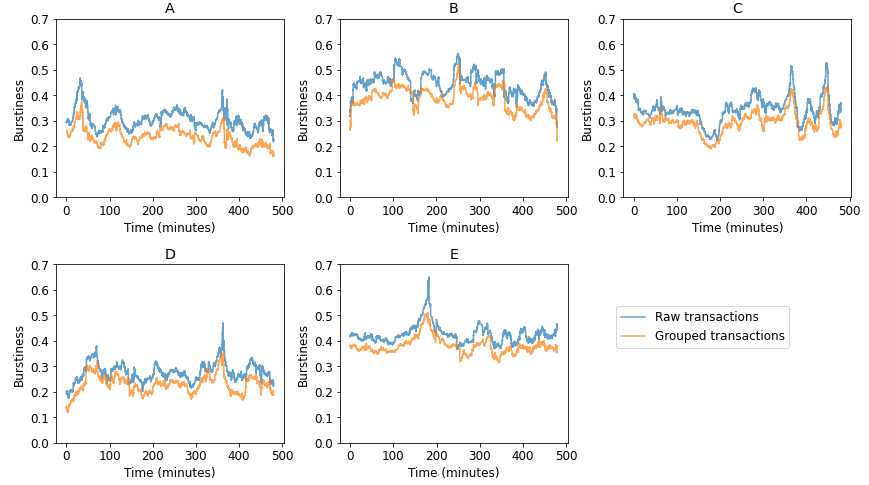}
    \caption{Rolling burstiness in windows containing 200 transactions, of all transactions for the instruments considered, both for the raw transaction timestamps and grouped timestamps.}
    \label{fig:raw_grouped_bursts}
\end{figure}

It is worth noting that a key property of these systems is the presence of `split' trades. It is common practice to split large trades into a series of smaller ones \cite{chou_09}, as it allows securities to be traded easily without being constrained by insufficient liquidity, and can make them more eligible for rapid execution. Split trades appear in transaction reports in bursts of transactions occuring at the same time or very close together. In order to understand the extent to which the observed burstiness in these systems is explained by trade splitting, we apply our methods both to the observed timestamps, and also to timestamps grouped at an edge level, considering transactions as part of the same order if the transactions in a group have inter-trade times of less than 1 second. Figure \ref{fig:raw_grouped_bursts} shows the rolling burstiness produced as in figure \ref{fig:rolling_network_burstiness} for all transactions, both for the observed transactions, and for transactions grouped by searching for bursts of transactions which have an inter-transaction time of less than 1 second. We can see that grouping the transactions produces a sequence with a slightly lower burstiness, however does not entirely remove the burstiness of these sequences. 

\begin{figure}
\centering
\begin{subfigure}{.4\textwidth}
  \centering
  \includegraphics[width=\linewidth]{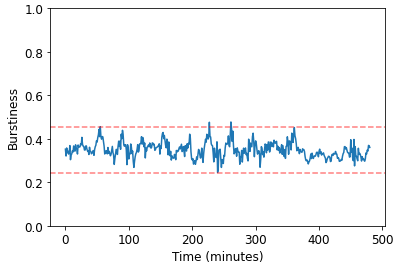}
\caption{Burstiness} 
 \label{fig:timeaveraged_burstiness}
 \end{subfigure}
\begin{subfigure}{.4\textwidth}
  \centering
  \includegraphics[width=\linewidth]{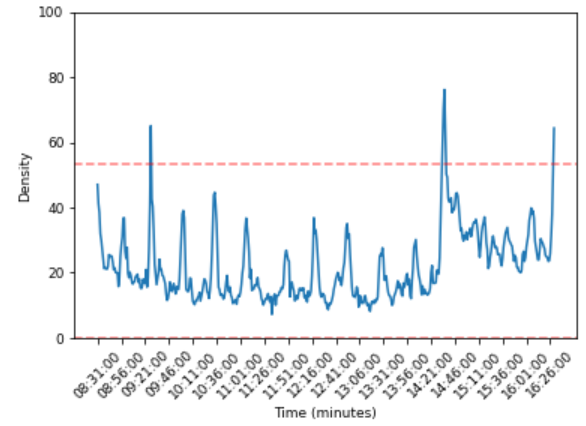}
    \caption{Density}
  \label{fig:timeaveraged_density}
    \end{subfigure}%
\caption{Rolling burstiness and density, averaged across four different days for one of the datasets considered. The dashed lines represent 3 standard deviations from the mean of the time series. The burstiness is computed with a window of 200 transactions, and the density over a 10 minute time window.}
\end{figure}

In addition to the single day explorations of these datasets, we also looked to see if there were any consistent patterns across multiple days in the burstiness, to understand whether the burstiness observed is affected by regular events e.g. different markets opening. Figures \ref{fig:timeaveraged_burstiness} and \ref{fig:timeaveraged_density} show the rolling burstiness and density respectively for one of the datasets, averaged over four different trading days. We can see that although there are a few significant peaks in the burstiness occurring around the middle of the day, the density shows significantly more prevalent peaks representing variations in market activity. Further to this, there is a clearly observable rise in the density in the later third of the day, which is not reflected in the burstiness suggesting that the burstiness can be treated as a noisy but fixed quantity across a day's trading.

% \begin{figure}
%     \centering
%     \includegraphics[width=\linewidth]{timeaveraged_burstiness.PNG}
%     \caption{Rolling burstiness over a window of 200 transactions, averaged across four different days for one of the datasets considered. The dashed lines represent 3 standard deviations from the mean of the time series.}
%     \label{fig:timeaveraged_burstiness}
% \end{figure}

% \begin{figure}
%     \centering
%     \includegraphics[width=\linewidth]{timeaveraged_density.PNG}
%     \caption{Rolling density over a time window of 300 seconds, averaged across four different days for one of the datasets considered. The dashed lines represent 3 standard deviations from the mean of the time series.}
%     \label{fig:timeaveraged_density}
% \end{figure}

\begin{figure}
    \centering
    \includegraphics[width=\linewidth]{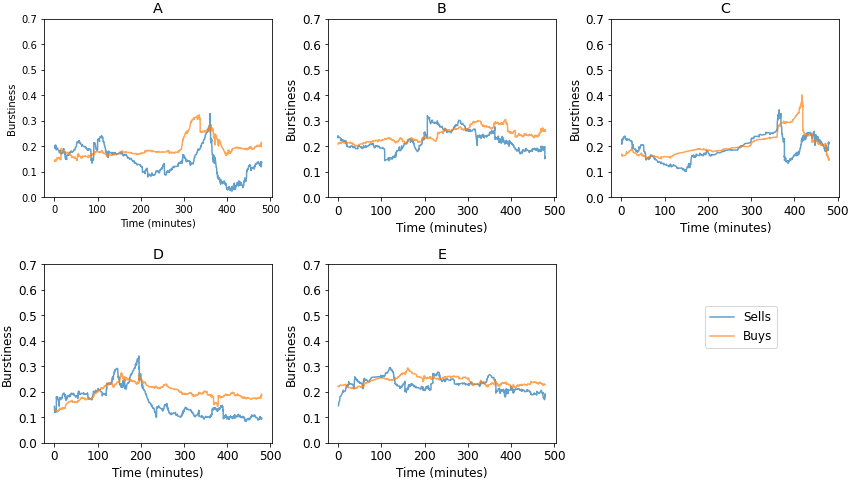}
    \caption{Rolling burstiness over a window of 200 transactions for the 5 different instruments, considering transaction sequences of buys and sells separately for transactions executed through the main clearing house.}
    \label{fig:bursts_buys_sells}
\end{figure}

Finally, we note that in the case of the subset of trades executed through a single clearing house, we can alternatively consider the transaction sequences of buys and sells separately. Figure \ref{fig:bursts_buys_sells} shows the rolling burstinesses of the buys and sells separately for trades executed through the most active clearing house. We can see from these that the burstinesses of the buys and sells show a similar level of burstiness, although there are periods for certain instruments where the two transaction sequences differ in their burstinesses. Visually, there also appears to be some influence of higher periods of burstiness of buys influencing a higher burstiness of sells.

To summarise, we observe in all 5 datasets burstiness appears to be an inherent property of the transaction sequences, that isn't solely caused by order splitting, or constrained to a particular type of trading activity. Across all transactions, the burstiness shows stability across time even when the underlying transaction density varies, however at the level of sequences between individual participants, the burstiness varies significantly with some participant pairs showing non-bursty behaviour. To try to take advantage of the bursty properties of the Hawkes process in the following sections we consider a number of different approaches to modelling transaction sequences across the various subsets of the instruments considered, making use of both univariate and multivariate Hawkes processes to establish the extent to which self- and mutual exciting behaviour could be responsible for the burstiness we observe.

\subsection{Generative models for systems with multiple hubs}
When considering transactions for a single instrument that is traded on multiple venues, through a single venue for which trades are cleared by multiple clearing houses, or considering off exchange transactions, we consider these systems as dynamic networks, for which the generative process can be modelled as a multivariate Hawkes process at the edge level. This allows both self excitation at the level of each edge, such that the presence of a transaction between two counterparties increases the likelihood of them transacting in the future, and mutual excitation, whereby the presence of a transaction between two counterparties increases the likelihood of transactions between other counterparties. 
In order to understand the prevalence of self- and mutual excitation, we also consider two alternative applications of the Hawkes process - first, considering transactions generated according to a single univariate process, and second, considering transactions generated according to univariate Hawkes processes at the level of each individual edge. 
We further compare these models to models using Poisson processes, considering both a single Poisson process for all transactions, and individual Poisson processes for each edge. 
For the scenario where we use a univariate Hawkes process to model the transaction sequence as a whole, our model consists of two parts - timestamp generation and edge selection, as a univariate Hawkes process does not inherently produce a cross sectional structure so this needs to be simulated separately.

\subsubsection{Univariate Hawkes processes}
\label{sec:univariate}
\begin{figure}
    \centering
    \includegraphics[width = \linewidth]{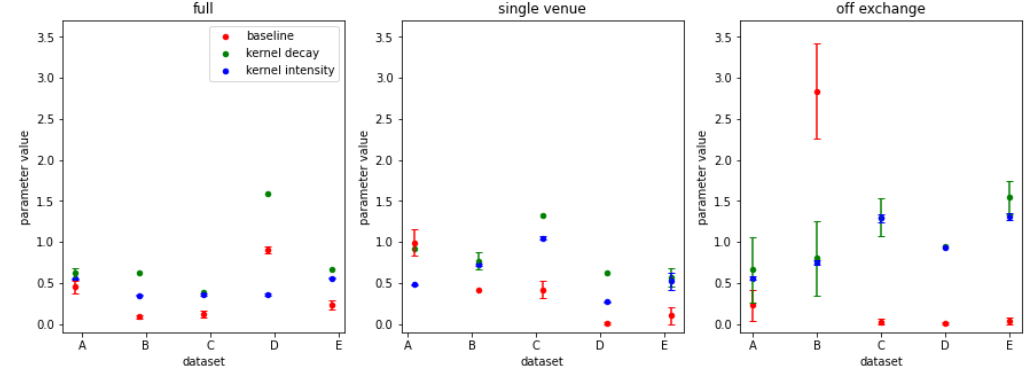}
    \caption{Parameter values for the 5 different datasets, across the three subsets considered. Error bars (where present) represent the standard error calculated by taking the inverse of the Hessian. The standard error is not calculated when the Hessian is non-invertible.}
    \label{fig:univ_params}
\end{figure}

 The results of using a Maximum Likelihood approach to fit the three parameters of a univariate process to the full transaction sequence are shown in tables \ref{tab:param_vals} and \ref{tab:burst_res}, with the former reporting the estimated parameters and the latter the burstiness of simulated sequences from the fitted model. Considering the parameter values, there is a reasonable amount of variation across the different datasets for all three subsets considered, as demonstrated in figure \ref{fig:univ_params}. For the baseline intensity and kernel decay parameters, there is no consistent difference in the parameter value for the different data subsets across all of the datasets considered, however the Kernel intensity appears to show the largest values for the off venue subset, and smaller values for the full dataset. Revisiting our earlier observation that the burstiness is highest for the full dataset and lowest for off venue transactions, this appears contradictory, suggesting that the model is not able to accurately estimate the parameters in the univariate case.
\begin{table}
    \centering
    \begin{tabular}{|c|c|c|c|}
    \hline
    Data & Baseline int. & Kernel int. & Kernel dec. \\
    \hline
    A (full) & $0.45 \pm$ 0.08 & $0.55 \pm 0.01$ & $0.62 \pm 0.06$\\
    A (single venue) & $0.99 \pm 0.16 $ &$0.48 \pm 0.01$  & $0.92 \pm $ NA \\
    A (off exchange) & $0.23 \pm 0.19$ & $0.56 \pm 0.02$ & $0.66 \pm 0.40 $ \\
    B (full) & $0.09 \pm 0.02 $ & $0.35 \pm 0.01$ & $0.63 \pm NA$ \\
    B (single venue) & $0.41 \pm NA$ & $0.72 \pm 0.01$ & $0.77 \pm 0.10 $ \\
    B (off exchange) & $2.84 \pm 0.58$ & $0.75 \pm 0.03$ & $0.80 \pm 0.45 $ \\
    C (full) & $0.12 \pm 0.04$ & $0.36 \pm 0.01$ & $0.38 \pm NA$ \\
    C (single venue) & $0.42 \pm 0.10$ & $1.05 \pm 0.02$ & $1.33 \pm NA$ \\
    C (off exchange) & $0.03 \pm 0.03$ & $1.29\pm 0.05$ & $1.30 \pm 0.23$ \\
    D (full) & $0.90 \pm 0.04$ & $0.36 \pm 0.01$ & $1.59 \pm NA$ \\
    D (single venue) & $0.01 \pm 0.01$  & $0.27 \pm 0.01$ & $0.63 \pm NA$ \\
    D (off exchange) & $0.01 \pm 0.01$ & $0.93 \pm NA$ & $0.94 \pm NA$\\
    E (full) & $0.23 \pm 0.06$ & $0.56 \pm 0.01$ & $0.66 \pm NA$ \\
    E (single venue) & $0.10 \pm 0.10$ & $0.52 \pm 0.10$ & $0.57 \pm 0.11$ \\
    E (off exchange) & $0.04 \pm 0.04$ & $1.31 \pm 0.04$ & $1.54 \pm 0.20$ \\
       \hline
    \end{tabular}
    \vspace{0.1cm}
    \caption{Hawkes parameters estimated using a Maximum Likelihood approach for the chosen datasets, across the different subsystems considered as dynamic networks}
    \label{tab:param_vals}
\end{table}

\begin{table}
    \centering
    \begin{tabular}{|c|c|c|c|}
    \hline
    Data & Real data & Poisson & Hawkes \\%& transaction density \\
    \hline
    A (full) & 0.40 & 0 (-0.01, 0.01) & 0.30 (0.27, 0.33)  \\
    A (single venue) & 0.30 &0 (-0.02, 0.02) & 0.13 (0.11,0.14) \\
    A (off exchange) & 0.29 & 0 (-0.02, 0.02) & 0.40 (0.34, 0.46) \\
    B (full) & 0.54 & 0 (-0.02, 0.02) & 0.27 (0.24, 0.29)  \\
    B (single venue) & 0.39 & 0 (-0.01, 0.01) & 0.41 (0.35, 0.46) \\
    B (off exchange) & 0.31  & 0 (-0.02, 0.02) & 0.08 (0.06, 0.11) \\
    C (full) & 0.44 & 0 (-0.01,0.01) & 0.51 (0.48, 0.56) \\
    C (single venue) & 0.32 & 0 (-0.01, 0.01) & 0.37 (0.34, 0.40) \\
    C (off exchange) & 0.32 & 0 (-0.02, 0.02) & 0.78 (0.48, 0.94) \\
    D (full) & 0.34 & 0 (-0.01,0.01) & 0.07 (0.06, 0.08)  \\
    D (single venue) & 0.26 & 0 (-0.02, 0.02) & 0.21 (0.01, 0.39) \\
    D (off exchange) & 0.19 & 0 (-0.02, 0.02) & 0.75 (0.37, 0.96) \\
    E (full) & 0.55 & 0 (-0.02, 0.02) & 0.41 (0.38, 0.43) \\
    E (single venue) & 0.45 & 0 (-0.02, 0.02) & 0.54 (0.48, 0.60) \\
    E (off exchange) & 0.27 & 0 (-0.02, 0.02) & 0.53 (0.43, 0.64) \\
      \hline
    \end{tabular}
    \vspace{0.1cm}
    \caption{Burstiness of the timestamps generated by Hawkes and Poisson processes, in comparison to the real timestamps}
    \label{tab:burst_res}
\end{table}

\begin{figure}
    \centering
    \includegraphics[width = \linewidth]{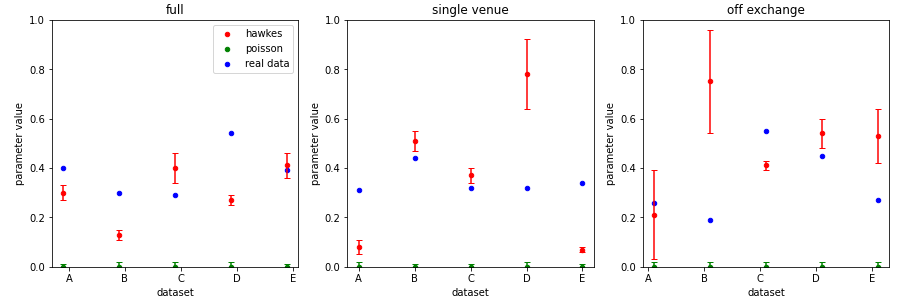}
    \caption{Burstiness of the simulated timestamps using the parameters in table 
    \ref{fig:univ_params}, for the 5 different datasets across the different subsets, in comparison to the burstiness of the real data and a univariate Poisson process fitted to the real data.}
    \label{fig:burst_uni}
\end{figure}

Table \ref{tab:burst_res} and figure \ref{fig:burst_uni} show the burstiness of sequences generated by Hawkes processes using the parameters in table \ref{tab:param_vals}. Although the Hawkes process produces sequences with a positive burstiness, and are in most cases closer to the real burstiness than a Poisson process fitted to the same data, the 95\% confidence intervals obtained when running the simulations 1000 times do not encapsulate the real burstiness for the majority of the datasets, so we can conclude that the univariate Hawkes process is not able to produce sequences with the same level of burstiness as the real data. This supports our observations of counterintuitive parameter values in that the univariate Hawkes process is not fully capturing the behaviour of these systems, suggesting that mutual excitation could be a relevant generative property of transaction sequences across the different subsets.

\begin{figure}
    \centering
    \includegraphics[width = \linewidth]{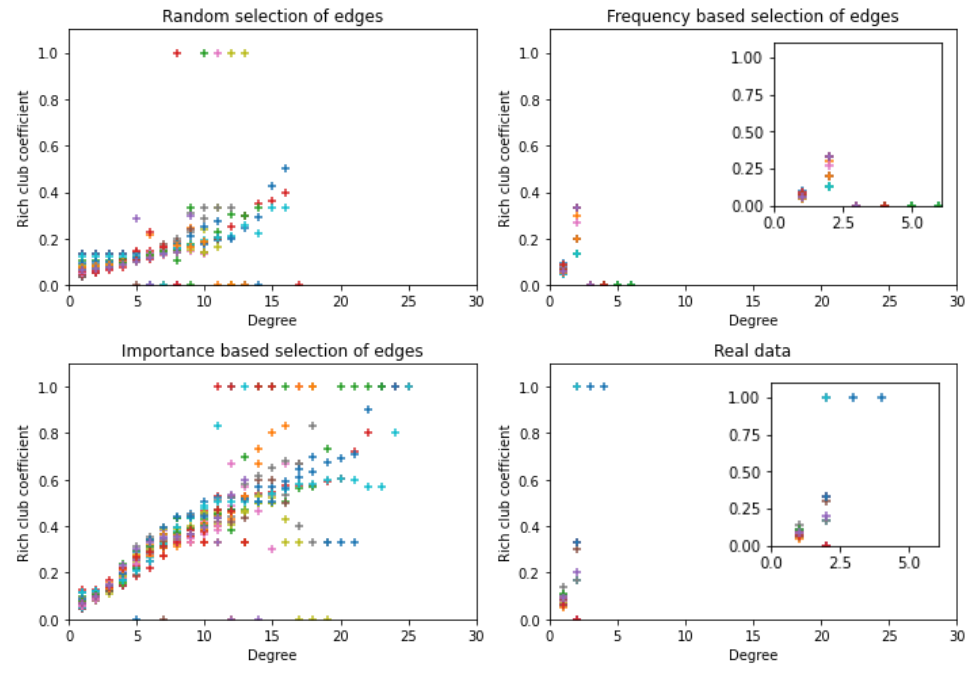}
    \caption{Rich club coefficients for the degrees of nodes in the network for the FTSE-A instrument. The colours represent snapshot networks aggregated every 10 minutes. Inset plots are presented for the frequency and real data with a reduced x axis scale.}
    \label{fig:rich_clubs}
\end{figure}

As explained above, when using a univariate Hawkes process, we consider three different methods to select the edge to change at each time step. To evaluate the performance of these models, we compare their ability to reproduce several properties of the real data. 
The first property we consider is the rich club coefficients at different node degrees. 
Figure \ref{fig:rich_clubs} the rich club coefficients at different node degrees for snapshots of networks every 10 minutes, for all transactions for instrument A for illustration. We see that for the real data, the nodes with the largest degree have very low rich club indices, which is driven by the presence of disconnected star subnetworks containing a single hub node (in all cases a CCP) connected to a number of other nodes that are only connected to the hub. We see that this property is also reproduced by frequency based edge selection, however is not for random selection or importance based selection. 

Table \ref{tab:ks_rich} shows the p-values of a two dimensional, two sided Kolmogorov-Smirnov test applied to each of the results of each of the different edge selection methods, in comparison to the real data. Here, small p-values mean that the two samples are significantly different, with the null hypothesis being that the two distributions are the same. We can see that this null hypothesis is rejected for all of the datasets with random or importance based selection, so these methods are not able to reproduce this property of the real data. However, we see that for frequency based selection, considering the full datasets, four of the datasets show a p-value of $>0.05$, meaning that for these the null hypothesis is not rejected and the distributions are not significantly different from the rich club distributions of the real data. For the dataset for which the null hypothesis is rejected for frequency based edge selection, the p-values are still much higher than for the random and importance based selection, suggesting some level of similarity between these distributions and the rich club distribution of the real data. Similar results are seen for the single venue subsets, where three of the datasets have p-values $>0.05$, however for off-exchange trading the null hypothesis is rejected at $5\%$ confidence, suggesting that transactions executed off exchange are less dependent on the prevalence of historical trades between the two counterparties that the frequency based selection models.

\begin{table}
    \centering
    \begin{tabular}{|c|c|c|c|}
    \hline
    Data & Random & Frequency & Importance \\%& transaction density \\
    \hline
    A (full) & $2.99 \times 10^{-20}$ & 0.38 & $1.55\times 10^{-20}$ \\
    A (single venue) & $1.88\times 10^{-20}$ & 0.73 & $6.12\times 10^{-44}$\\
    A (off exchange) & $3.01\times 10^{-6}$ & 0.06 & $9.00 \times 10^{-7}$ \\
    B (full) & $4.20 \times 10^{-20}$ & 0.36 & $3.86\times 10^{-22}$ \\
    B (single venue) & $9.39\times 10^{-18}$ & 0.030 & $2.44\times 10^{-39}$\\
    B (off exchange) & $8.63\times 10^{-7}$ & 0.01 & $2.85\times 10^{-6}$\\
    C (full) & $1.74\times 10^{-33}$ & 0.41 & $7.09\times 10^{-21}$ \\
    C (single venue) &  $3.53\times 10^{-24}$ & 0.001 & $8.61\times 10^{-37}$\\
    C (off exchange) & $8.21\times 10^{-5}$ & 0.01 & $3.31\times 10^{-5}$ \\
    D (full) & $2.41\times 10^{-31}$ & 0.035 & $2.91\times 10^{-32}$ \\
    D (single venue) & $2.44\times 10^-{13}$ & 0.35 & $4.70\times 10^{-32}$\\
    D (off exchange) & $7.04\times 10^{-8}$ & 0.04 & $3.59\times 10^{-8}$ \\
    E (full) & $4.87\times 10^{-36}$ & 0.018 & $1.76\times 10^{-37}$ \\
    E (single venue) & $1.67\times 10^{-24}$ & 0.18 & $1.00\times 10^{-36}$\\
    E (off exchange) & $9.50\times 10^{-9}$ & $1.02\times 10^{-4}$ & $5.57\times 10^{-8}$ \\
      \hline
    \end{tabular}
    \vspace{0.1cm}
    \caption{p-values for a two dimensional two sided Kolmogorov-Smirnov test comparing the distributions of the Rich Club index over different degrees for the three different edge selection methods in comparison to the real data, for the 5 datasets considered.}
    \label{tab:ks_rich}
\end{table}

\begin{figure}
    \centering
    \includegraphics[width=\linewidth]{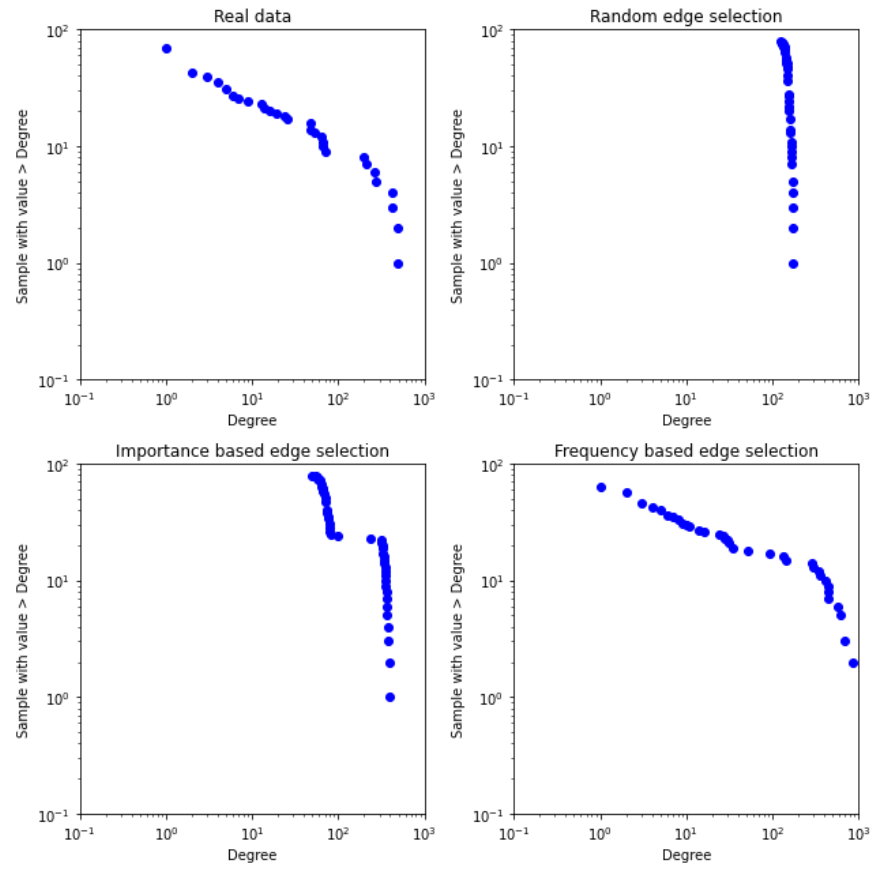}
    \caption{Cumulative degree distributions for the full networks computed over the entire time period, plotted on a log-log scale}
    \label{fig:degree_dists}
\end{figure}

The second property we consider is the aggregated degree distributions. Figure \ref{fig:degree_dists} shows an example of the degree distributions for the three different edge selection methods along with the real data. Here we see that the real data and frequency based edge selection data show similar shaped degree distributions, with a small number of nodes with very high degree, but the majority of nodes with very low degree. The random and importance based edge selection methods instead produce networks which have a large number of nodes with a reasonably high degree.  Table \ref{tab:ks_deg} shows the p-values of a two dimensional, two sided Kolmogorov-Smirnov test applied to each of the degree distributions of the simulated data, in comparison to the real data. We can see that the null hypothesis of the samples having the same distribution is not rejected for all of the datasets for the frequency based edge selection, but we observe very small p-values for the other two methods confirming that these methods are not able to reproduce the degree distribution of the real data. 

\begin{table}
    \centering
    \begin{tabular}{|c|c|c|c|}
    \hline
    Data & Random & Frequency & Importance \\%& transaction density \\
    \hline
    A (full) & $2.78 \times 10^{-48}$ & $0.10$ & $4.27\times 10^{-97}$\\
    A (single venue) &  $7.12\times 10^{-61}$ & 0.15 & $3.96\times 10^{-167}$ \\
    A (off exchange) & $3.73 \times 10^{-27}$ & $0.01$ & $1.20 \times 10^{-30}$ \\
    B (full) & $4.24 \times 10^{-71}$ & 0.99 &$2.74\times 10^{-146}$ \\
    B (single venue) & $1.79\times 10^{-63}$ & 0.15 &  $5.46\times 10^{-163}$ \\
    B (off exchange) & $4.53 \times 10^{-27}$ & $4.03 \times 10^{-9}$ & $1.28\times 10^{-28}$ \\
    C (full) & $5.43 \times 10^{-40}$& 0.06 &$9.53 \times 10^{-78}$\\
    C (single venue) & $1.86\times 10^{-34}$  & 0.12 &  $1.74 \times 10^{-93}$ \\
    C (off exchange) & $3.40 \times 10^{-18}$ & $1.75 \times 10^{-5}$ & $2.29 \times 10^{-25}$ \\
    D (full) & $2.20 \times 10^{-65}$ & 0.99 & $2.59 \times 10^{-114}$ \\
    D (single venue) & $4.79\times 10^{-72}$ & $0.45$ & $3.76\times 10^{-176}$ \\
    D (off exchange) & $6.25 \times 10^{-34}$ & $0.02$& $6.74 \times 10^{-42}$ \\
    E (full) & $2.41 \times 10^{-91}$ & 0.99 & $2.27 \times 10^{-135}$\\
    E (single venue) & $4.71\times 10^{-98}$ & 0.24 & $2.54\times 10^{-184}$\\
    E (off exchange) & $3.73 \times 10^{-30}$ & $0.99$ & $4.82\times 10^{-32}$\\
      \hline
    \end{tabular}
    \vspace{0.1cm}
    \caption{p-values for a two dimensional two sided Kolmogorov-Smirnov test comparing the degree distributions for the three different edge selection methods in comparison to the real data, for the 5 datasets considered.}
    \label{tab:ks_deg}
\end{table}

\begin{figure}
    \centering
    \includegraphics[width=\linewidth]{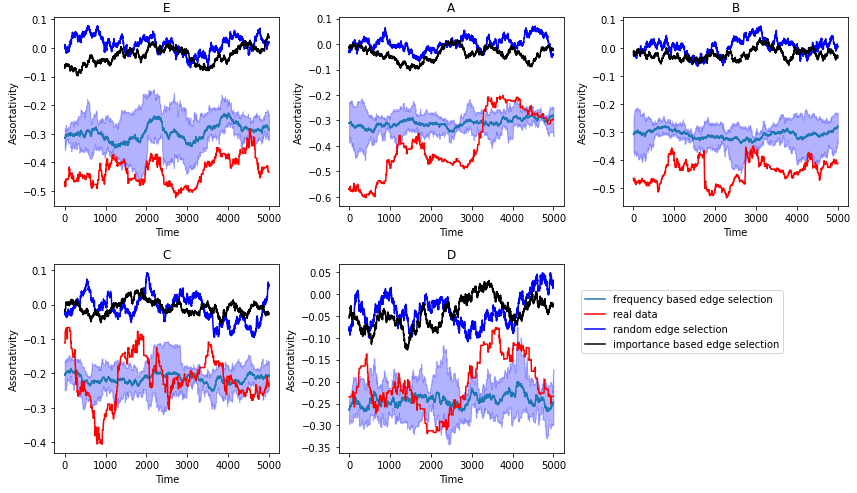}
    \caption{Assortativity computed across time in rolling windows of 200 transactions for the 5 different datasets, presented for the three methods of edge selection along with the real data. In the case of the frequency based method, the shaded blue area corresponds to 95\% Confidence Intervals. }
    \label{fig:assortativity}
\end{figure}

Next we consider the assortativity observed in both the aggregated network across the entire time period, and also in rolling windows across time. Figure \ref{fig:assortativity} shows the assortativity across time computed over a rolling window for the 5 different datasets for the 3 different methods of edge selection. We can clearly see that the frequency based edge selection method produces an assortativity that is similar to that of the real data, whereas the random and importance based selection are unable to reproduce the assortativity of the real data. The blue bands represent $95\%$ confidence intervals for the assortativity for the frequency based method, and although the real data assortativity falls within the confidence intervals some of the time, this is generally less than $95\%$ of the time so we cannot conclude that the frequency based edge selection method is producing sequences with the same assortativity as the real data.

\begin{figure}
    \centering
    \includegraphics[width=\linewidth]{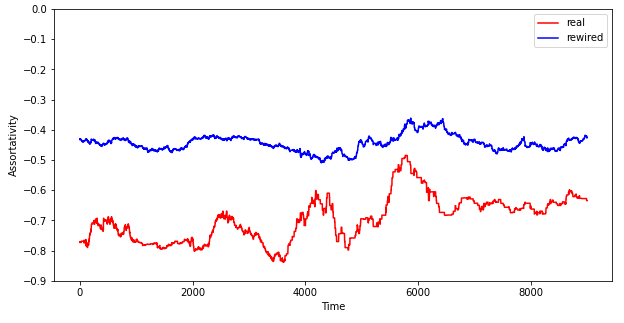}
    \caption{Assortativity computed across time in rolling windows of 200 transactions for dataset A, in comparison to the assortativity computed for dataset A with edge rewiring applied $N$ times, with $N$ the number of edges in the network}
    \label{fig:assortativity_rewire}
\end{figure}

Care must be taken in interpreting the degree distribution and assortativity results for the case of frequency based selection, as the method of sampling the edges to change probabilistically based on their frequency essentially fixes the degree distribution for a large enough sample size when the degree distribution is temporally stable. The values that the assortativity can take when the degree distribution is fixed will be constrained, since the degree distribution constrains the configuration of the network, which places bounds on the values that the assortativity can take. To understand the extent to which this explains the observed assortativity, we conduct an experiment in which the edges of the network are randomly re-wired to destroy any correlations between neighbours while preserving the degree distribution. The results of the assortativity across a rolling window a tenth of the length of the transaction sequence for both the real network and the rewired is shown in figure \ref{fig:assortativity_rewire}. We can see from this that although some of the assortativity is preserved when rewiring, not all of the observed assortativity in the real data is explained by the constraints imposed by the selection method. 

\begin{figure}
    \centering
    \includegraphics[width=\linewidth]{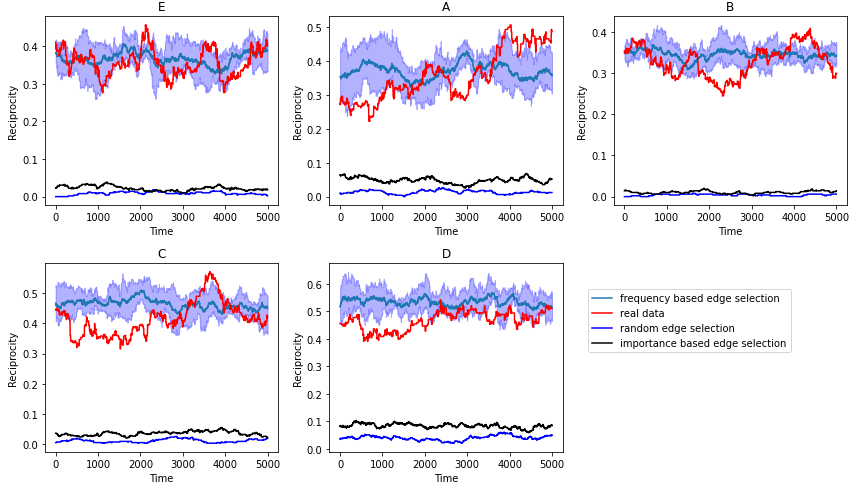}
    \caption{Reciprocity computed in rolling windows of 200 transactions for the 5 different datasets, presented for the three methods of edge selection along with the real data.}
    \label{fig:reciprocity}
\end{figure}

Finally, we consider reciprocity again both in the aggregated network across the entire time period and in rolling windows. Figure \ref{fig:reciprocity} shows the reciprocity across time computed over a rolling window for the 5 different datasets and 3 methods of edge selection. Again we clearly see that the frequency based method produces a similar reciprocity to the real data, but the other two methods are unable to reproduce the reciprocity. The reciprocity of the real data also falls within the $95\%$ confidence intervals of the frequency based selection method more often than in the case of the assortativity, however this is still less than $95\%$ of the time for four of the five datasets. 

\subsubsection{Edge level Hawkes processes}
We now consider making use of Hawkes processes to generate transaction sequences at the edge level. When generating transactions in this way, we no longer need to select the edge to change at each timestamp. We compare both generating edge transaction sequences for the individual edges independently using univariate Hawkes processes, as well as using a multivariate Hawkes process which allows the edges to be mutually exciting. As a benchmark, we also compare these to using edge level Poisson processes. 
It is worth noting that one of the major drawbacks of using an edge level Hawkes process, is that the majority of edges in the network only appear once in a single day, so when considering edge level Hawkes processes, we restrict our networks to edges which appear more than 10 times over the day considered. This roughly halves the number of counterparties for all of the datasets considered. 

We make use of the ADM4 method \cite{zhou_2013} which uses a fixed, given kernel decay for all kernels. This means that we need to estimate the decay using a brute force approach, in which we fit the multivariate Hawkes process for a range of decay values, and observe the burstinesses of the individual edges for the resulting simulations. We then choose the best decay value for each edge, and take the median of these decays as the optimal burstiness for the final simulation.

Table \ref{tab:bursts_multiv} shows the burstiness of the simulated transactions for the different datasets across the different subsets, in comparison to the real burstinesses and univariate and multivariate Poisson processes. We see here that the univariate approach has the best performance, showing the true burstiness falling within the 95\% confidence intervals in all cases for the off exchange data subsets, and 3 of the 5 datasets for the full and single venue subsets. This suggests that much of the burstiness of the real data can be attributed to self-excitation at the edge level, or in other words, that the presence of a trading relationship between two counterparties increases the likelihood of future trading relationships. The multivariate approach shows the true burstiness falling outside the confidence intervals for the majority of the trials, however the method itself produces results with tighter confidence intervals around the mean, suggesting a higher level of stability in the resulting simulations. 

\begin{table}
    \centering
    \begin{tabular}{|c|c|c|c|c|c|}
    \hline
     & True & univ. Poiss. & multiv. poiss. & Univ. HP & Multiv. HP \\%& transaction density \\
    \hline 
    A (full) & 0.45 & 0 (0,0.01) & 0 (-0.06, 0.07) & $\mathbf{0.48 (0.41,0.56)}$ & 0.35 (0.33, 0.40)  \\
    A (single venue) & 0.21 & 0 (-0.01,0.01) & 0 (-0.05, 0.05) & $\mathbf{0.15 (0.07, 0.25)}$ & $\mathbf{0.24 (0.21, 0.26)}$\\
    A (off exchange) & 0.18 & 0 (-0.01, 0.01) & 0 (-0.06, 0.07) & $\mathbf{0.18 (0.09, 0.30)}$ & 0.26 (0.21, 0.30) \\
    B (full) & 0.60 & 0 (0, 0.01) & 0 (-0.06, 0.07) & 0.43 (0.34, 0.58) & 0.41 (0.39, 0.46) \\
    B (single venue) & 0.16 &  0 (0, 0.01)& 0 (-0.07, 0.07) & $\mathbf{0.16 (0.08, 0.26)}$ & 0.28 (0.25, 0.31) \\
    B (off exchange) & 0.10 & 0 (-0.01, 0.01) & 0 (-0.06, 0.07) & $\mathbf{0.12 (0.02, 0.23)}$ & 0.17 (0.13, 0.21) \\
    C (full) & 0.54 & 0 (0, 0.01) & 0 (-0.06, 0.07) & $\mathbf{0.54 (0.49, 0.60)}$ & $\mathbf{0.44 (0.35, 0.62)}$  \\
    C (single venue) & 0.37  & 0 (-0.01, 0.01) & 0 (-0.06, 0.07)  & 0.20 (0.10, 0.33) & 0.19 (0.16, 0.22) \\
    C (off exchange) & 0.23 & 0 (0, 0.01) & 0 (-0.05, 0.05) & $\mathbf{0.16 (0.07, 0.26)}$ & $\mathbf{0.24 (0.20, 0.29)}$ \\
    D (full) & 0.43 & 0 (-0.01, 0.01) & 0 (-0.07, 0.07)  & 0.26 (0.20, 0.37) & 0.58 (0.53, 0.66)  \\
    D (single venue) & 0.34 & 0 (0, 0.01) & 0 (-0.05, 0.05) & 0.22 (0.13, 0.32) & 0.20 (0.18, 0.22) \\
    D (off exchange) & 0.20 & 0 (-0.01, 0.01) & 0 (-0.06, 0.06) & $\mathbf{0.15 (0.07, 0.23)}$ & $\mathbf{0.18 (0.15, 0.22)}$ \\
    E (full) & 0.56 & 0 (0, 0.01) & 0 (-0.06, 0.06) & $\mathbf{0.55 (0.50, 0.59)}$ & 0.34 (0.31, 0.38)  \\
    E (single venue) & 0.18 & 0 (-0.01, 0.01)& 0 (-0.07, 0.07) & $\mathbf{0.13 (0.05, 0.23)}$ & 0.28 (0.25, 0.31) \\
    E (off exchange) & 0.10 & 0 (0, 0.01) & 0 (-0.06, 0.07) & $\mathbf{0.17 (0.07, 0.28)}$ & 0.14 (0.12, 0.17) \\
      \hline
    \end{tabular}
    \vspace{0.1cm}
    \caption{Burstiness of the timestamps generated by edge level Hawkes and Poisson processes, in comparison to the real timestamps. Results shown in bold are the cases where the true burstiness falls within the 95\% confidence intervals of the simulation results.}
    \label{tab:bursts_multiv}
\end{table}

\begin{figure}
    \centering
    \includegraphics[width=\linewidth]{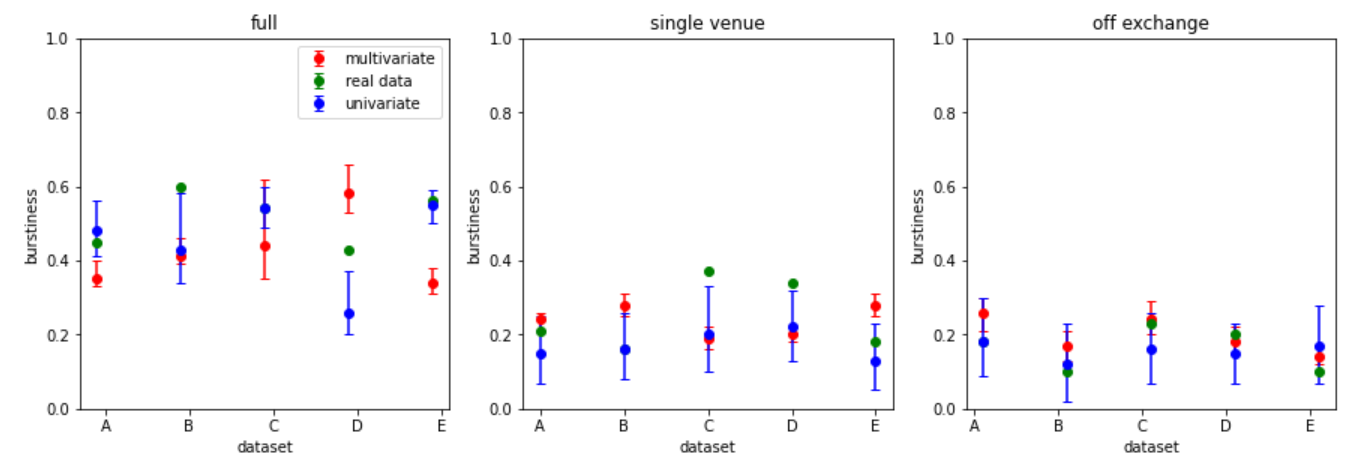}
    \caption{Burstiness of the simulated timestamps for both univariate and multivariate edge level Hawkes, for the 5 different datasets across the different subsets, in comparison to the burstiness of the real data. }
    \label{fig:burst_multivariate}
\end{figure}

\begin{figure}
    \centering
    \includegraphics[width=\linewidth]{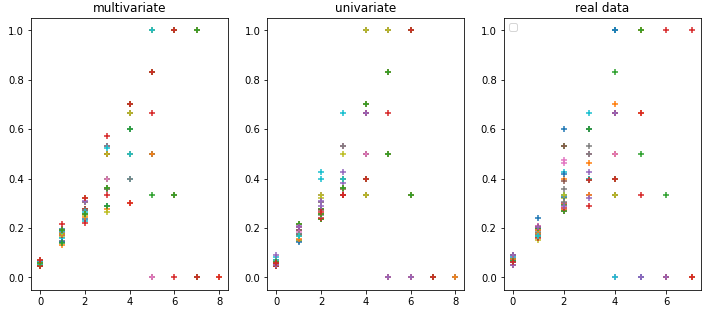}
    \caption{Distributions of rich club coefficients for the degrees of nodes in networks from snapshots aggregated hourly.}
    \label{fig:richclub_multivariate}
\end{figure}

In contrast to the univariate case, when considering edge level models we no longer require a method to select which counterparties transact at each timestep. In this next section, we restrict ourselves to considering the full datasets as the network structure diminishes when considering these smaller subsets following the removal of infrequent counterparty relationships. In our considerations of edge selection methods in section \ref{sec:univariate}, we considered the rich club and degree distributions, as well as reciprocity and assortativity. Considering these same properties, starting with the rich club coefficient distributions, figure \ref{fig:richclub_multivariate} demonstrates the similarities in both simulation methods to the real data. This is confirmed in all cases using the same Kolmogorov-Smirnov test as before, for which the null hypothesis of differences between the distributions is rejected for both models, for all datasets.

% \begin{figure}
%     \centering
%     \includegraphics[width=\linewidth]{degree_distributions_multi.PNG}
%     \caption{Degree distribution of networks from snapshots aggregated hourly.}
%     \label{fig:degree_dist_multivariate}
% \end{figure}

\begin{figure}
    \centering
    \includegraphics[width=\linewidth]{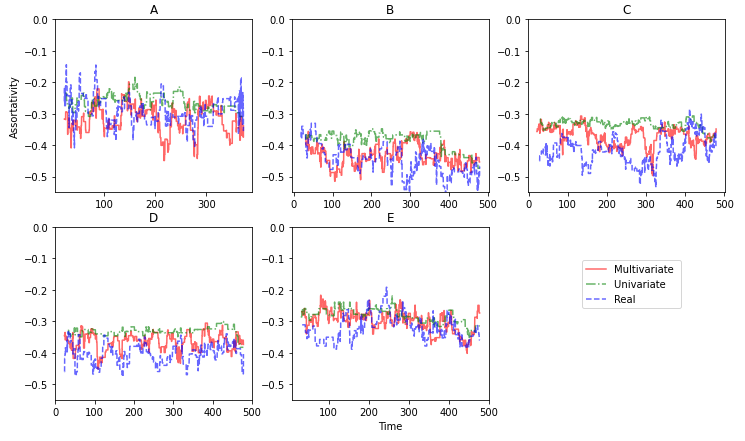}
    \caption{Assortativity computed across time in rolling windows of 200 transactions for the 5 different datasets, presented for the two different methods of generating edge level transaction sequences, in comparison to the real data.}
    \label{fig:assortativity_multi}
\end{figure}

\begin{figure}
    \centering
    \includegraphics[width=\linewidth]{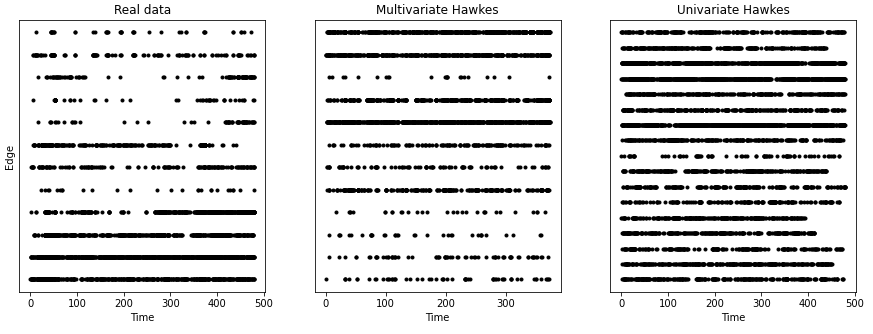}
    \caption{Point plots for the individual edges for the two methods of generating edge level transaction sequences, in comparison to the real data}
    \label{fig:seq_plot_multi}
\end{figure}

\begin{figure}
    \centering
    \includegraphics[width=\linewidth]{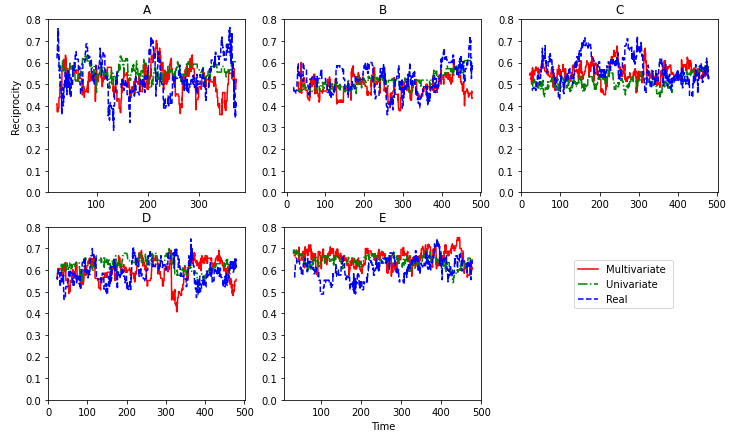}
    \caption{Reciprocity computed across time in rolling windows of 200 transactions for the 5 different datasets, presented for the two different methods of generating edge level transaction sequences, in comparison to the real data.}
    \label{fig:reciprocity_multi}
\end{figure}

Figures \ref{fig:assortativity_multi} and \ref{fig:reciprocity_multi}  show the assortativity and reciprocity computed over rolling windows for both methods of applying Hawkes processes to generate edge level transaction sequences. We can see here that both the assortativity and reciprocity of the sequences generated using a multivariate Hawkes are similar in value and variance to the real process, however those generated using the univariate Hawkes process show near a slightly lower variance, albeit at a similar level to the multivariate approach and the real data. This low variance can be understood by visually inspecting the transaction sequences, as in figure \ref{fig:seq_plot_multi}, in which we plot the point processes of the individual edges. We can see that the univariate approach produces a much lower prevalence of low density transaction sequences than observed for the real data. This is likely a result of poor fit of the Hawkes processes at the individual level, producing similar parameter values for very different sequences, which corresponds with our observations of wider confidence intervals for the univariate burstiness, suggesting a lower stability of this model in comparison to the multivariate approach.
Although there is a visually high similarity of both reciprocity and assortativity for the multivariate Hawkes process sequences in comparison to the real data, in making use of a Levene test to assess whether the sequences have equal variance, and also making use of a one-sided t-test to assess whether the mean of the difference between the simulated and actual reciprocity and assortativity is significantly different from 0, in all cases the p-values were <0.5, meaning that we cannot accept the null hypothesis in either case and validate the similarity between the sequences.

To summarise the performance of both the transaction level and edge level approaches for systems with multiple hubs, we have presented the burstinesses of all of the Hawkes approaches in a single table in table \ref{tab:all_bursts}. Here we can see that the univariate edge level Hawkes process shows the strongest overall performance, although this is heavily dependent on the wider confidence intervals of this model.  
\begin{table}
    \centering
    \begin{tabular}{|c|c|c|c|c|}
    \hline
     & True & univ. HP & univ. edge HP & multiv. HP \\%& transaction density \\
    \hline 
    A (full) & 0.45 &  0.30 (0.27, 0.33) & $\mathbf{0.48 (0.41,0.56)}$ & 0.35 (0.33, 0.40)  \\
    A (single venue) & 0.21 &  0.13 (0.11, 0.14) & $\mathbf{0.15 (0.07, 0.25)}$ & $\mathbf{0.24 (0.21, 0.26)}$\\
    A (off exchange) & 0.18 &  0.40 (0.34, 0.46) & $\mathbf{0.18 (0.09, 0.30)}$ & 0.26 (0.21, 0.30) \\
    B (full) & 0.60  & 0.27 (0.24, 0.29) & 0.43 (0.34, 0.58) & 0.41 (0.39, 0.46) \\
    B (single venue) & 0.16 & 0.41 (0.35, 0.46) & $\mathbf{0.16 (0.08, 0.26)}$ & 0.28 (0.25, 0.31) \\
    B (off exchange) & 0.10 & $\mathbf{0.08 (0.06, 0.11)}$ & $\mathbf{0.12 (0.02, 0.23)}$ & 0.17 (0.13, 0.21) \\
    C (full) & 0.54  & $\mathbf{0.51 (0.48, 0.56)}$ & $\mathbf{0.54 (0.49, 0.60)}$ & $\mathbf{0.44 (0.35, 0.62)}$  \\
    C (single venue) & 0.37  & $\mathbf{0.37 (0.34, 0.40)}$  & 0.20 (0.10, 0.33) & 0.19 (0.16, 0.22) \\
    C (off exchange) & 0.23 & 0.78 (0.48, 0.94) & $\mathbf{0.16 (0.07, 0.26)}$ & $\mathbf{0.24 (0.20, 0.29)}$ \\
    D (full) & 0.43 & 0.07 (0.06, 0.08)  & 0.26 (0.20, 0.37) & 0.58 (0.53, 0.66)  \\
    D (single venue) & 0.34 & $\mathbf{0.21 (0.01, 0.39)}$ & 0.22 (0.13, 0.32) & 0.20 (0.18, 0.22) \\
    D (off exchange) & 0.20 & 0.75 (0.37, 0.96) & $\mathbf{0.15 (0.07, 0.23)}$ & $\mathbf{0.18 (0.15, 0.22)}$ \\
    E (full) & 0.56 & 0.41 (0.38, 0.43) & $\mathbf{0.55 (0.50, 0.59)}$ & 0.34 (0.31, 0.38)  \\
    E (single venue) & 0.18 & 0.54 (0.48, 0.60) & $\mathbf{0.13 (0.05, 0.23)}$ & 0.28 (0.25, 0.31) \\
    E (off exchange) & 0.10 & 0.53 (0.43, 0.64) & $\mathbf{0.17 (0.07, 0.28)}$ & 0.14 (0.12, 0.17) \\
      \hline
    \end{tabular}
    \vspace{0.1cm}
    \caption{Burstiness of the timestamps generated by Hawkes processes at the level of all transactions and edge level, in comparison to the real timestamps. Results shown in bold are the cases where the true burstiness falls within the 95\% confidence intervals of the simulation results.}
    \label{tab:all_bursts}
\end{table}

\subsection{Bivariate Hawkes model for buy and sell sequences}
When considering trades occurring through a central clearing party, from a networks perspective we have a network consisting of a single star, and we are limited in the insights we can gain into the behaviour of the system by modelling it as a dynamic network. Instead, we re-frame the modelling of the system to consider the buy and sell trade executions as separate sequences, and allow for mutual excitation between the buys and sells. This is similar in approach to the recently published methods in 
As before, we make use of Maximum Likelihood estimation to fit the parameters of the Hawkes process to the transaction sequences. Since we are considering a bivariate Hawkes process, there are 8 parameters to estimate with a $2 \times 2$ matrix for the adjacency $\mathbf{\alpha}$ two decay parameters $\mathbf{\beta}$, and two baseline intensity parameters $\mu_i$. %Instead of estimating all parameters, we propose an approach to simplify the parameter space, by estimating the decay parameters as the decay of a univariate Hawkes process fitted to the buy and sell sequences individually. As these univariate decays will capture the self-excitatory properties of the sequences, but not the mutual excitation between the sequences, we apply a scaling factor to the decays to produce a burstiness closest to the real burstiness. TODO - use R method to estimate all params for the bivariate process. 

\begin{figure}
    \centering
    \includegraphics[width=\linewidth]{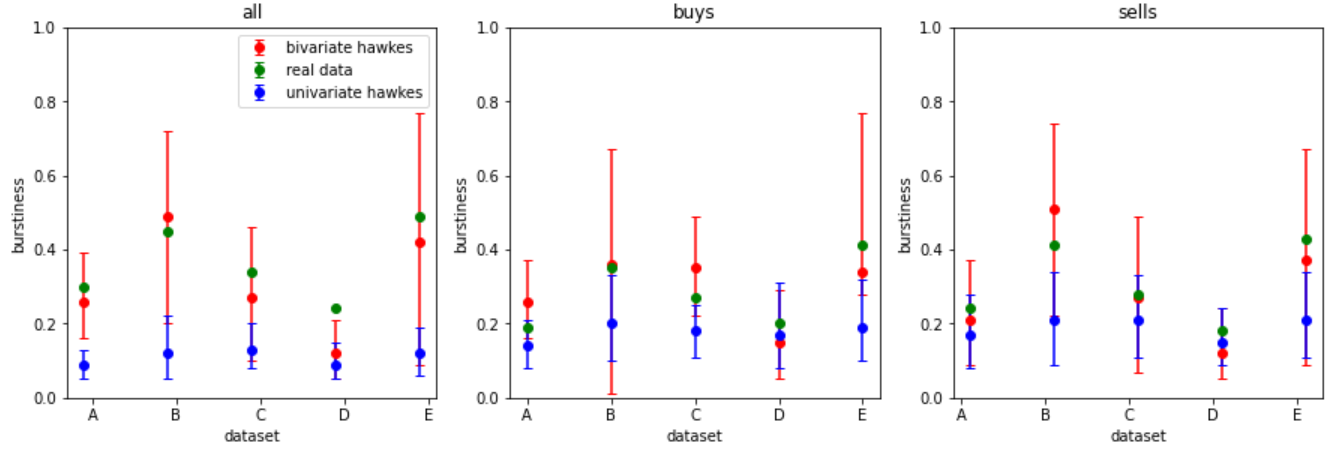}
    \caption{Burstiness of the simulated timestamps using a bivariate Hawkes, for the 5 different datasets considering trades through a single central clearing house, in comparison to the burstiness of the real data. }
    \label{fig:burst_biv}
\end{figure}

The estimated parameters for the bivariate Hawkes model are shown in table \ref{tab:param_vals_biv}, and the resulting burstiness of the overall sequences, as well as the buy and sell sequences separately, are shown in table \ref{tab:burst_biv} and figure \ref{fig:burst_biv}. We see here that all datasets show similar value ranges for the parameters, and that we do not observe higher or lower parameter values for buys in comparison to sells consistently across the different datasets. Considering the burstiness, we see that in the majority of cases, the true burstiness falls within the confidence intervals of the simulations for both the full transaction sequences as well as the buys and sells themselves. For the case of dataset D, which shows a true burstiness overall outside of the confidence intervals of the simulation, if we look back to figure \ref{fig:bursts_buys_sells}, we see that the burstiness of the sells drops during the latter part of the day, which explains why our model is unable to reproduce the overall burstiness. We can also compare to fitting Hawkes processes to the buys and sells separately, also shown in table \ref{tab:burst_biv}, to help us understand how much the observed burstiness is driven by self excitation as opposed to mutual excitation. We see here that although the confidence intervals of the buys and sells contains the true burstiness in many cases, the simulated transaction sequences as a whole have a consistently lower burstiness than the real transaction sequences. This means that some of the burstiness of this system is likely to be driven by mutual excitation between buy trades and sell trades.

\begin{table}
    \centering
    \begin{tabular}{|c|c|c|c|}
    \hline
    Data & Baseline int. & Kernel int. & Kernel dec. \\
    \hline
 & & & \\
    A  & $[0.24, 0.11]$  & 
 $\begin{bmatrix}
   0.57 & 0.42 \\
   0.15 & 0.10 \\
  \end{bmatrix} $ 
 & $[1.23, 0.24]$\\
 & & & \\
    A (SE) & $[0.05, 0.07]$ &    $\begin{bmatrix}
   0.04 & 0.03 \\
   0.07 & 0.03 \\
  \end{bmatrix}$ & $[0.03, 0.08]$ \\
   & & & \\
    B  & $[0.03, 0.01]$  & 
 $\begin{bmatrix}
   0.15 & 0.0 \\
   0.21 & 0.58 \\
  \end{bmatrix}  $
 & $[0.15, 0.76]$\\
  & & & \\
    B (SE) & $[0.03, 0.01]$ &    $\begin{bmatrix}
   0.03 & 0.01 \\
   0.04 & 0.05 \\
  \end{bmatrix}$ & $[0.03, 0.06]$ \\
   & & & \\
    C  & $[0.29, 0.02]$  & 
$ \begin{bmatrix}
   0.59 & 0.01 \\
   0.05 & 0.20 \\
  \end{bmatrix} $ 
 & $[0.67, 0.25]$\\
  & & & \\
    C (SE) & $[0.07, 0.02]$ &    $\begin{bmatrix}
   NA & 0.03 \\
   0.09 & 0.21 \\
  \end{bmatrix}$ & $[NA, 0.30]$ \\
   & & & \\
    D  & $[0.20, 0.53]$  & 
 $\begin{bmatrix}
   0.21 & 0.07 \\
   0.05 & 0.27 \\
  \end{bmatrix} $ 
 & $[0.33, 0.37]$\\
  & & & \\
    D (SE) & $[0.13, 0.07]$ &    $\begin{bmatrix}
   0.04 & 0.03 \\
   0.02 & NA \\
  \end{bmatrix}$ & $[0.08, NA]$ \\
   & & & \\
    E  & $[0.00, 0.04]$  & 
 $\begin{bmatrix}
   0.42 & 0.01 \\
   0.27 & 1.07 \\
  \end{bmatrix} $ 
 & $[0.43, 1.67]$\\
  & & & \\
    E (SE) & $[0.00, 0.02]$ &    $\begin{bmatrix}
   0.04 & 0.02 \\
   0.05 & 0.10 \\
  \end{bmatrix}$ & $[0.04, 0.22]$ \\
   & & & \\
       \hline
    \end{tabular}
    \vspace{0.1cm}
    \caption{Hawkes parameters estimated using a Maximum Likelihood approach for the chosen datasets, for a bivariate Hawkes process applied to buys and sells through a central clearing house}
    \label{tab:param_vals_biv}
\end{table}

\begin{table}
    \centering
    \begin{tabular}{|c|c|c|c|}
    \hline
    Data & Real data & Univariate & Bivariate \\%& transaction density \\
    \hline
    A (all) & 0.30 & 0.09 (0.05, 0.13) & 0.26 (0.16,0.39) \\
    A (buys) & 0.19 & 0.14 (0.08, 0.21) & 0.26 (0.16, 0.37) \\
    A (sells) & 0.24 & 0.17 (0.08, 0.28) & 0.21 (0.09, 0.37) \\
    B  & 0.45 & 0.12 (0.05, 0.22) & 0.49 (0.20, 0.72) \\
    B (buys) & 0.35 & 0.20 (0.10 0.33) & 0.36 (0.01, 0.67) \\
    B (sells) & 0.41 & 0.21 (0.09, 0.34) & 0.51 (0.22, 0.74) \\
    C  & 0.34 & 0.13 (0.08, 0.20) & 0.27 (0.10, 0.46) \\
    C (buys) & 0.27 & 0.18 (0.11, 0.25) & 0.35 (0.22, 0.49) \\
    C (sells) & 0.28 & 0.21 (0.11, 0.33) & 0.27 (0.07, 0.49) \\
    D  & 0.24 & 0.09 (0.05, 0.15) &  0.12 (0.05, 0.21) \\
    D (buys) & 0.20 & 0.17 (0.08, 0.31) &  0.15 (0.05, 0.29) \\
    D (sells) & 0.18 & 0.15 (0.09, 0.24) & 0.12 (0.05, 0.24)\\
    E  & 0.49 & 0.12 (0.06, 0.19) & 0.42 (0.09, 0.77) \\
    E (buys) & 0.41 & 0.19 (0.10, 0.32) & 0.34 (0.28, 0.77) \\
    E (sells) & 0.43 & 0.21 (0.11, 0.34) & 0.37 (0.09, 0.67) \\
      \hline
    \end{tabular}
    \vspace{0.1cm}
    \caption{Burstiness of timestamps generated with a bivariate Hawkes process.}
    \label{tab:burst_biv}
\end{table}

\section{Discussion}
The application of various formulations of Hawkes processes to transaction sequences for FTSE100 stocks gives us a novel insight into the behaviours of these markets, and shows the promise of Hawkes processes as generative models for transaction sequences. As these markets are highly heterogeneous, with many different methods of trading giving rise to different behaviours, we have considered our methods in application to all transactions available for a single day, as well as only those transactions that are executed on the dominant exchange, and only those that are executed off exchange. 

When considering the univariate approach, we observed counter-intuitive parameter values and although the model was able to reproduce sequences with a burstiness closer to the real data than the benchmark of a Poisson process, in most cases this model underestimated the burstiness, suggesting that self-excitation at the transaction level does not accurately represent the true generative process of these transaction sequence. In selecting which pair of counterparties should transact at each point in the transaction sequence, selecting based on historical frequency produced cross-sectional properties that were the most similar to those of the real data, suggesting that historical counterparty relationships influences the probability that a pair of counterparties will transact in the future. 

When considering the edge level approaches, the approach making use of a univariate process for each edge was able to reproduce the burstiness of the real dataset much more often than the mutlivariate approach could, however, the multivariate approach showed tighter $95\%$ confidence intervals for the burstiness, suggesting that this model was more stable. The better performance of the univariate edge model is in agreement with the strong performance of the frequency based edge selection, since both suggest that historical counterparty relationships are influential in the presence of a future relationship. However, the multivariate approach performed much better than the univariate edge model in reproducing the cross sectional properties of the transaction networks, suggesting that improvements to this model, for example allowing the kernel decay to vary across the different edges, would be a promising avenue for further research.

Finally, we observed strong performance in applying a bivariate model to buys and sells for trades occurring through a single central clearing counterparty, with the burstiness reproduced in the majority of cases. We also observed a better performance of the bivariate approach in comparison to a univariate hawkes for buys and sells separately suggesting that mutual excitation between the buys and the sells contributes to the overall burstiness. Further exploration into the performance of edge level modelling in conjunction with the bivariate model would be an interesting avenue of further exploration. 

Since we are proposing generative methods of transaction sequences, it is worth discussing the potential uses of simulated transaction sequences. In an ideal world in which transaction reporting contains no errors, duplicate reports or nuances which cause deviations from the generative model, these methods would be valuable in anomaly detection, in flagging transactions which deviate too far from the model, or alternatively systems which show differing aggregate properties, allowing consideration of how these properties constitute risk. For example, we can see that all 5 systems considered in this paper show similar burstinesses across time, similar degree distributions, and similar reciprocities and assortativities across time. The presence of burstiness shows the market response to activity, which drives price formation, and the stability of prices is integral to the good function of markets as a whole. The degree distributions we observe demonstrate fat tails in which the majority of participants have a low degree with a small number of hub nodes with very large degrees, usually central clearing houses which play a large part in mitigating the risks of highly interconnected networks and the propagation of risk. This is also the dominant property of the rich club distributions, and the presence of disassortativity, in which nodes tend to connect to nodes that they are dissimilar to, or in other words play a different role within these networks. The presence of reciprocity suggests that normal functioning markets rely on the presence of well established relationships between participants - this will usually be due to the presence of the clearing member- clearing house relationships, whereas if we observe a shift where market participants 'shop around', this could be suggestive of a lack of good service and a potential stability risk. 
It is also worth noting that a number of problems encountered in finance involve extreme class imbalances, for example fraud detection. A generative model for transactions could be used as an oversampling method to generate artificial data, to help alleviate class imbalance, as is explored in Hung et. al. \cite{Hung_2019}, who make use of Generative Adversarial Networks to assist on the classification of credit card fraudulent transactions, which are highly imbalanced with only $0.17\%$ of transactions of the positive (fraudulent) class.  

Although we have considered application of different models to different subsets of the trading ecosystems studied, there are several trading mechanisms that we have not yet explored in terms of their influence on the systems' properties. For example, in markets such as these which are heavily dominated by trades which are centrally cleared, in order to provide clearing members with increased opportunities to net their orders and to give a reduction in outstanding gross exposures in the system, interoperability agreements exist which allow CCPs to link between counterparties allowing cross-system execution of transactions. These links introduce a direct channel of contagion between these critical nodes in the financial system, so in order for our methods to fully capture the macroprudential risks in these systems, these links would need to be included. Further to this, we have considered application of Hawkes processes to trade executions only, and further research into how the dynamics of the order book influences the generative processes of these executions would be useful to explore in addition to the other potential avenues suggested.

\section*{Availability of data and materials}
All five datasets studied in this paper were extracted from a dataset of transaction reports collected by the FCA under MIFID II regulations. The datasets were used under agreement from the data owners at the Financial Conduct Authority for the current study, and are not publicly available. 

An implementation of the methods referenced in this paper in a mixture of Python and R can be found at \cite{my_code}. 
%Different levels of burstiness characterised by whether traders are splitting orders? %https://www.researchgate.net/publication/228318347_Modelling_Trades-Through_in_a_Limit_Order_Book_Using_Hawkes_Processes 
%In order to grasp the clustering of trades-though, we compute the mean of
%the distribution of waiting times between two consecutive trades-through, and we
%compare it with the mean waiting time between one trade (of any kind) and the
%next trade-through.
% given the empirical evidence of burstiness... 

% It appears that we observe very large variations in the results of the numeri-
% cal maximization of the likelihood. However, whatever the absolute size of the
% parameters, it is clear that the cross-excitation effect, i.e. the excitation of trades-
% through of a given side by the occurrence of trades-through on the opposite side,
% is much weaker than the self-excitation effect, which translates the clustering of
% trades-through on a given side.
%Royal society open access or EPJ B
\bibliographystyle{unsrt}
\bibliography{bibliography.bib}
\end{document}